\documentclass[useAMS,usenatbib]{mn2e}
\usepackage{graphicx}
\newcommand{\lre}{$\log R_{\rm e}$}
\newcommand{\re}{$R_{\rm e}$}

\newcommand{\sn}{$n$}
\newcommand{\mie}{$<\! \mu\! >_{\rm e}$}
\newcommand{\ls}{$\log \sigma_0$}
\newcommand{\olre}{$\overline{\log R_e}$}
\newcommand{\omie}{$\overline{\log <\! \mu\! >_e}$}
\newcommand{\ols}{$\overline{\log \sigma_0}$}
\newcommand{\dagk}{$\delta a_{g->K}$}
\newcommand{\dbgk}{$\delta b_{g->K}$}
\newcommand{\be}{\begin{equation}}
\newcommand{\ee}{\end{equation}}
\newcommand{\bea}{\begin{eqnarray}}
\newcommand{\eea}{\end{eqnarray}}

\title[Environmental Dependence of the FP]{SPIDER -- III.   Environmental  Dependence  of  the
  Fundamental Plane of Early-type Galaxies}
\author[F. La Barbera et al.]{F. La Barbera$^{1}$\thanks{E-mail: labarber@na.astro.it}, P.A.A. Lopes$^{2}$, R.R. de Carvalho$^{3}$, I.G. de la Rosa$^{4,5}$, A.A. Berlind$^{6}$
 \\
$^1$INAF -- Osservatorio Astronomico di Capodimonte, Napoli, Italy \\
$^2$Observat\'orio do Valongo/UFRJ, Rio de Janeiro, Brazil\\
$^3$Instituto Nacional de Pesquisas Espaciais/MCT, S. J. dos Campos, Brazil\\
$^4$Instituto de Astrofisica de Canarias (IAC), E-38200 La Laguna, Tenerife, Spain\\
$^5$Depto. de Astrofisica, Universidad de La Laguna (ULL), E-38206 La Laguna, Tenerife, Spain\\
$^6$Vanderbilt University Dept. of Physics and Astronomy, Nashville, USA}

\begin{document}

\date{Accepted; Received}

\pagerange{\pageref{firstpage}--\pageref{lastpage}} \pubyear{2010}

\maketitle

\label{firstpage}

\begin{abstract}
We analyse the Fundamental  Plane (FP) relation of $39,993$ early-type
galaxies  (ETGs)  in  the  optical  (griz) and  $5,080$  ETGs  in  the
Near-Infrared  (YJHK) wavebands,  forming an  optical$+$NIR  sample of
$4,589$ galaxies.  We focus on the analysis of the FP as a function of
the environment where galaxies reside. We characterise the environment
using the largest group catalogue, based on 3D data, generated from SDSS
at low  redshift ($z < 0.1$).   We find that the  intercept $``c''$ of
the FP decreases  smoothly from high to low  density regions, implying
that  galaxies at  low  density have  on  average lower  mass-to-light
ratios  than  their   high-density  counterparts.   The  $``c''$  also
decreases as a function of  the mean characteristic mass of the parent
galaxy group. However, this trend is weak and completely accounted for
by the variation of $``c''$  with local density.  The variation of the
FP offset  is the  same in  all wavebands, implying  that ETGs  at low
density have  younger luminosity-weighted ages  than cluster galaxies,
consistent with  the expectations of semi-analytical  models of galaxy
formation.   We measure an  age variation  of $\sim  0.048$~dex ($\sim
11\%$)  per decade  of  local  galaxy density.   This  implies an  age
difference of about $32 \%$ ($\sim  3 \, Gyr$) between galaxies in the
regions of  highest density  and the field.   We find  the metallicity
decreasing, at $\sim  2$~$\sigma$, from low to high  density.  We also
find $2.5 \, \sigma$ evidence that  the variation in age per decade of
local density augments,  up to a factor of  two, for galaxies residing
in massive relative to poor  groups.  The velocity dispersion slope of
the FP,  $``a''$, tends  to decrease with  local galaxy  density, with
galaxies in  groups having  smaller $``a''$ than  those in  the field,
independent of the waveband used to measure the structural parameters.
Environmental  effects (such  as tidal  stripping) may  elucidate this
result, producing  a steeper variation of  dark-matter fraction and/or
non-homology  along the  ETG's  sequence at  higher  density.  In  the
optical, the  surface brightness slope,  $``b''$, of the  FP increases
with local  galaxy density, being  larger for group relative  to field
galaxies.  The difference vanishes in  the NIR, as field galaxies show
a small ($\sim  2.5\%$), but significant increase of  $``b''$ from $g$
through  $K$,  while  group   galaxies  (particularly  those  in  rich
clusters) do not.   The trend of $``b''$ with  the environment results
from galaxies residing  in more massive clusters, since  for groups no
variation  of  $``b''$ with  local  density  is  detected. A  possible
explanation  for  these findings  is  that  the  variation of  stellar
population properties with  mass in ETGs is shallower  for galaxies at
high  density, resulting from  tidal stripping  and quenching  of star
formation in galaxies  falling into the group's potential  well. We do
not detect  any dependence of the  FP coefficients on  the presence of
substructures in parent galaxy groups.
\end{abstract}


\section{Introduction}
\label{sec:intro}
A robust  prediction of hierarchical  theories of galaxy  formation is
that the environment  plays an important role in  shaping the galaxian
properties.   From the observational  viewpoint, this  is successfully
confirmed  by the  existence, for  instance, of  the  well established
morphology-density  relation~\citep{Dressler:80},   for  which  denser
environments  are  preferably populated  by  galaxies with  early-type
morphology in contrast to the late-type dominated field regions.

The role of environment in shaping  the properties of ETGs -- the most
massive galaxies in the local Universe -- is still an open question in
our understanding  of galaxy formation and evolution.   Studies of the
colour--magnitude (hereafter CM) relation in the low-redshift-Universe
have found  no significant  difference in the  average colour  of ETGs
among  high-  and low-density  environments,  implying  either a  tiny
difference or a strong anti-correlated variation of stellar population
properties,       i.e.       age      and       metallicity,      with
environment~\citep{Bern:03b,   Hogg:04,    Haines:06}.    Even   small
differences  in the  star formation  history of  galaxies  residing in
different environments would be magnified as one approaches the galaxy
formation epoch.  Several studies  have found no remarkable difference
in the properties of cluster  and field galaxies at high redshift. For
instance, \citet{Koo:05a,  Koo:05b} found that both  field and cluster
ETGs   have  very   similar   CM  relations   at   redshift  $z   \sim
0.8$. \citet{Cool:06} also  reported that the offset and  slope of the
CM relation of  field and cluster galaxies are  fully consistent up to
redshift  $z  \sim 0.4$.   \citet{Pannella:09}  found  that ETGs  have
similar  characteristic  ages,  independent  of the  environment  they
belong to.  An opposite picture  comes out of spectroscopic studies of
ETGs  at redshift  $z \sim  0$.   Many authors  have reported  younger
luminosity-weighted ages for field  relative to cluster galaxies, with
the     age    difference     amounting    to     $\sim    1$--$2$~Gyr
(e.g.~\citealt{GLC:92, Trager:00,  Kunt:02, TF:02, Thomas:05, BERN:06,
  Clemens:09}).  Such difference seems  to exist also when considering
galaxies  residing in  high-density, low-velocity  dispersion systems,
such as the Hickson Compact Groups~\citep{deLaRosa:07}.

A even more puzzling issue is  the difference in metal content of ETGs
among  different environments,  as different  studies have  found that
field ETGs are more metal-rich~\citep{Thomas:05, deLaRosa:07, Kunt:02,
  Clemens:09}, as metal-rich  as~\citep{BERN:06, Annibali:07}, or even
more  metal-poor than  their cluster  counterparts~\citep{GALL:06}.  A
main feature  of the spectroscopic  investigations is that  the galaxy
spectra, and  hence the inferred stellar  population properties, refer
to the galaxy inner region,  typically inside one effective radius. On
the other  hand, ETGs are  know to possess internal  colour gradients,
with  their  central  part   being  redder  than  the  outskirts  {
  (see~\citealt{Wu:05,  Roche:09};  and  references therein)}.   These
gradients are mainly due  to metallicity~\citep{Pel:90}, with a small,
but  significant   contribution  from  age~\citep{LdC09,  Clemens:09}.
Moreover, as shown by~\citet{LaB:05},  colour gradients seem to depend
on  the  environment where  galaxies  reside,  hence complicating  the
environmental comparison of spectroscopic properties.

Due to its  small intrinsic dispersion (see e.g.~\citealt{Gargiulo:09,
  HB:09}; and references therein), the Fundamental Plane (FP) relation
of  ETGs~\citep{Dressler87,  George87}, i.e.   the  scaling law  involving
radius,  velocity dispersion,  and surface  brightness, is  a powerful
tool to measure their mass-to-light ratio (see e.g.~\citealt{vDF:96}).
{ This  results  from  interpreting  the   FP  as  a
  consequence of mass-to-light  ratio systematically varying with mass
  and the non-homology of  ETGs (see e.g.~\citealt{HjM95, CdC95, CL97,
    GrC97, BCC97, Tru04, BBT07,  TNR:09})}.  Several studies have used
the FP to  constrain the environmental variation, at  a given mass, of
galaxy  luminosity.   Even  in  this  case,  a  contradictory  picture
emerges.   ~\citet{BERN:06}   found  the  FP  relation   at  low-  and
high-density  to  have consistent  slopes  but  a significant  offset,
interpreting it as a difference of $\sim 1$~Gyr in the formation epoch
of  field and  cluster  ETGs.  On  the contrary,  \citet{Donofrio:08},
analysing galaxies in massive  nearby clusters, found also significant
variations of the slopes with local density.  High redshift studies of
the FP have reported a consistent evolution of the mass-to-light ratio
of  ETGs among  different  environments, implying  the same  formation
epoch  of field and  cluster galaxies~{  ~\citep{JCF:06, vDvM:06}}.
\citet{vW:05} also  found that the difference  in mass-to-light ratios
between field  and cluster galaxies  depends on galaxy mass,  with low
mass   systems  exhibiting  a   strong  differential   evolution  (see
also~\citealt{PDdC01, dSA:06}).

This work  is the  third paper  of a series  aimed to  investigate the
properties  of  bright  ($M_r<  -20$)   ETGs  as  a  function  of  the
environment   where   they   reside,  in   the   low-redshift-Universe
($0.05<z<0.1$).    The   Spheroid's   Panchromatic  Investigation   in
Different Environmental Regions (SPIDER; see La Barbera et al.  2010a,
hereafter paper I) is based on a sample of 39,993 ETGs with photometry
and  spectroscopy   available  from  SDSS,   and  Near-Infrared  (NIR)
photometry   available  from  the   UKIDSS-Large  Area   Survey  (LAS;
~\citealt{Law07}).   Here, we  define the  environment where  the ETGs
reside,  and investigate the  environmental dependence  of the  FP. To
this effect, we cross-correlate the SPIDER sample to the largest group
catalogue, based on  3D data, generated from SDSS at  low redshift ($z <
0.1$).  This  group catalogue contains  10,124 systems selected  using a
redshift-space friends-of-friends  (FoF) algorithm from  the SDSS data
release 7  (DR7, covering 9,380  square degrees of the  northern sky).
{ Notice  that we  chose not to  cross-correlate the  SPIDER sample
  with  existing SDSS-based  group catalogues,  as they  do  not cover
  either the  entire area  or the (low-)redshift  range of  the SPIDER
  survey.   For instance,  the maxBCG  catalogue is  a  volume limited
  cluster   sample  spanning   the   reshift  range   $0.10   <  z   <
  0.30$~\citep{Koester:07}, while the SPIDER  sample is at $z<0.1$. 
  The  C4 catalogue~\citep{MILLER:05} spans the redshift
  range of $0.02$ to $0.17$, but it covers only $2,600$ square degrees
  on the sky, being based on  SDSS-DR2. The FoF group/cluster catalogue we use
  in the present study has the main advantage of providing a complete
  sample of  groups and clusters  in the local Universe,  covering the
  large  sky area  of  SDSS-DR7.  This  allows  us} to  derive the  FP
relation in  a wide range of  environments using both  optical and NIR
data, and characterising the environments in terms of both $``local''$
(e.g.   galaxy density)  and  $``global''$ (e.g.   parent group  mass)
observables.  The  large sample size provided by  the SDSS+UKIDSS data
allows us not only to analyse the offset but also the slopes of the FP
as a function of environment.

The layout  of the paper is the  following. In Sec.~\ref{sec:samples},
we describe the samples of ETGs. Sec.~\ref{sec:group_catalog} presents
the catalogue of  galaxy groups we use to  define the environment, while
Sec.~\ref{sec:group_properties} describes  the properties measured for
each  group,  that  define   the  $``global''$  environment  of  ETGs.
Sec.~\ref{sec:environment} deals  with the measurement  of quantities,
such   as  local   galaxy   density,  that   define  the   $``local''$
environment. Sec.~\ref{sec:fp} introduces the FP relation, and the way
we measure  its slopes and intercept.   In Sec.~\ref{sec:offset_fp} we
derive the intercept of the FP in different wavebands as a function of
environment.  Sec.~\ref{sec:offset_sp} analyses  how the offset of the
FP constrains the difference  in stellar population content, i.e.  age
and    metallicity,   of    ETGs    among   different    environments.
Sec.~\ref{sec:slopes_fp} describes the dependence of the slopes of the
FP on local and global environment. In Sec.~\ref{sec:conc}, we discuss
the  main   results  of   this  work.   A   summary  is   provided  in
Sec.~\ref{sec:summ}. Throughout  the paper, we adopt  a cosmology with
$\Omega_{\rm m}=$  0.3, $\Omega_{\lambda}=$ 0.7,  and H$_0 =  75$ $\rm
km$ $s^{-1}$ Mpc$^{-1}$.

\section{Samples of ETGs}
\label{sec:samples}
The  SPIDER sample consists  of $39,993$  ETGs, with  available $griz$
photometry  and  spectroscopy  from  SDSS-DR6. Out  of  them,  $5,080$
galaxies have  also photometry available in the  $YJHK$ wavebands from
UKIDSS-LAS  (see paper  I).  All  galaxies have  two estimates  of the
central  velocity dispersion,  one  from SDSS-DR6  and an  alternative
estimate  obtained   by  fitting   SDSS  spectra  with   the  software
STARLIGHT~\citep{CID05}.  In  all wavebands, structural  parameters --
i.e.  the  effective radius, \re,  the mean surface  brightness within
that  radius, \mie,  and the  Sersic index,  \sn \,  -- have  been all
homogeneously measured by the software 2DPHOT~\citep{LBdC08}.

For the present work, we select two samples of ETGs.  First, we define
an optical, r-band sample of ETGs, consisting of all galaxies brighter
than $M_r=-20.55$,  and with $\chi^2_r<3$,  where $M_r$ is  the 2DPHOT
total galaxy magnitude  and $\chi^2_r$ is the reduced  $\chi^2$ of the
two-dimensional Sersic fitting in r-band (see paper I). This selection
leads  to  a volume  complete  sample  of  ETGs, with  better  quality
structural  parameters.  This  optical sample  consists  of $N=36,124$
ETGs.  We  also define  an optical+NIR sample,  by selecting  only the
galaxies  with photometry  available in  all wavebands,  brighter than
$M_r=-20.55$, and  with $\chi^2 <  3$ in all  wavebands~\footnote{ Our
  results are not affected by the choice of the $\chi^2$ cutoff value.
  Using  a value  of $2$  would not  change at  all  the environmental
  trends  of FP  coefficients.}.   This selection  leads  to a  volume
complete  sample  of  $4,589$   ETGs,  with  good  quality  structural
parameters from $g$ through $K$.

The dependence of the FP relation on the environment is first analysed
by   dividing   both   samples   into   field   and   group   galaxies
(Sec.~\ref{sec:field_group_samples}).   For  the  subsamples of  group
galaxies, the  role of  environment is also  analysed with  respect to
group properties, such  as mass, galaxy offset from  the group centre,
density  (measured  relative  to  other  group members),  as  well  as
substructure    (Sec.~\ref{sec:environment}).     As    detailed    in
Sec.~\ref{sec:field_group_samples},   the   r-band   sample  of   ETGs
($N=36,124$)  is  split into  two  subsamples  of  $16,717$ group  and
$11,824$  field   galaxies.   For  the  optical+NIR   sample  of  ETGs
($N=4,589$),  the subsamples of  group and  field galaxies  consist of
$2,187$ and $1,359$ ETGs, respectively.

\section{The group catalogue}
\label{sec:group_catalog}

\subsection {The updated FoF catalogue}
The updated FoF group catalogue  from SDSS contains $10,124$ systems and
is created as described in Berlind et al.~(2006), being the difference
the area used  ($9,380$ square degrees, compared to  the original area
of $3,495$  square degrees from DR3),  as the new version  is based on
the DR7. Besides the coordinates  and central redshift, the list gives
estimates  of  the total  number  of galaxies  in  the  group and  the
velocity dispersion.  However, we  decided to re-estimate  the central
redshift  of  the groups,  applying  the  gap technique  \citep{ada98,
  lop07,  lop09a} to  the central  (0.67 Mpc)  galaxies  (requiring at
least  three galaxies in  this analysis).  Doing that  we are  able to
re-derive   redshifts   for   $9,235$   ($90.4$\%)  systems   of   the
$``original''$  sample.   For these  groups  we  apply the  ``shifting
gapper''  technique \citep{fad96}  as  in \citet{lop09a}  in order  to
obtain  a list  of group  members  independent of  the FoF  algorithm.
These groups are then subjected  to a virial analysis analogous to the
one described  in \citet{gir98}, \citet{pop05,  pop07}, \citet{biv06},
\citet{lop09a}.   This  procedure   yields  estimates  of  $\sigma_P$,
$R_{500}$, $R_{200}$, $M_{500}$ and  $M_{200}$ for most of groups from
the FoF  sample. When  performing the virial  analysis we  require the
existence  of  at  least  five  galaxies  within  the  maximum  radius
considered.  Out of  the 9235 groups we are able  to measure the above
parameters  for   $8,083$,  i.e.   $79.2$\%  of  the   initial  sample
(comprising $10,124$ systems).

\begin{figure*}
\begin{center}
\leavevmode
\includegraphics[width=7.2in]{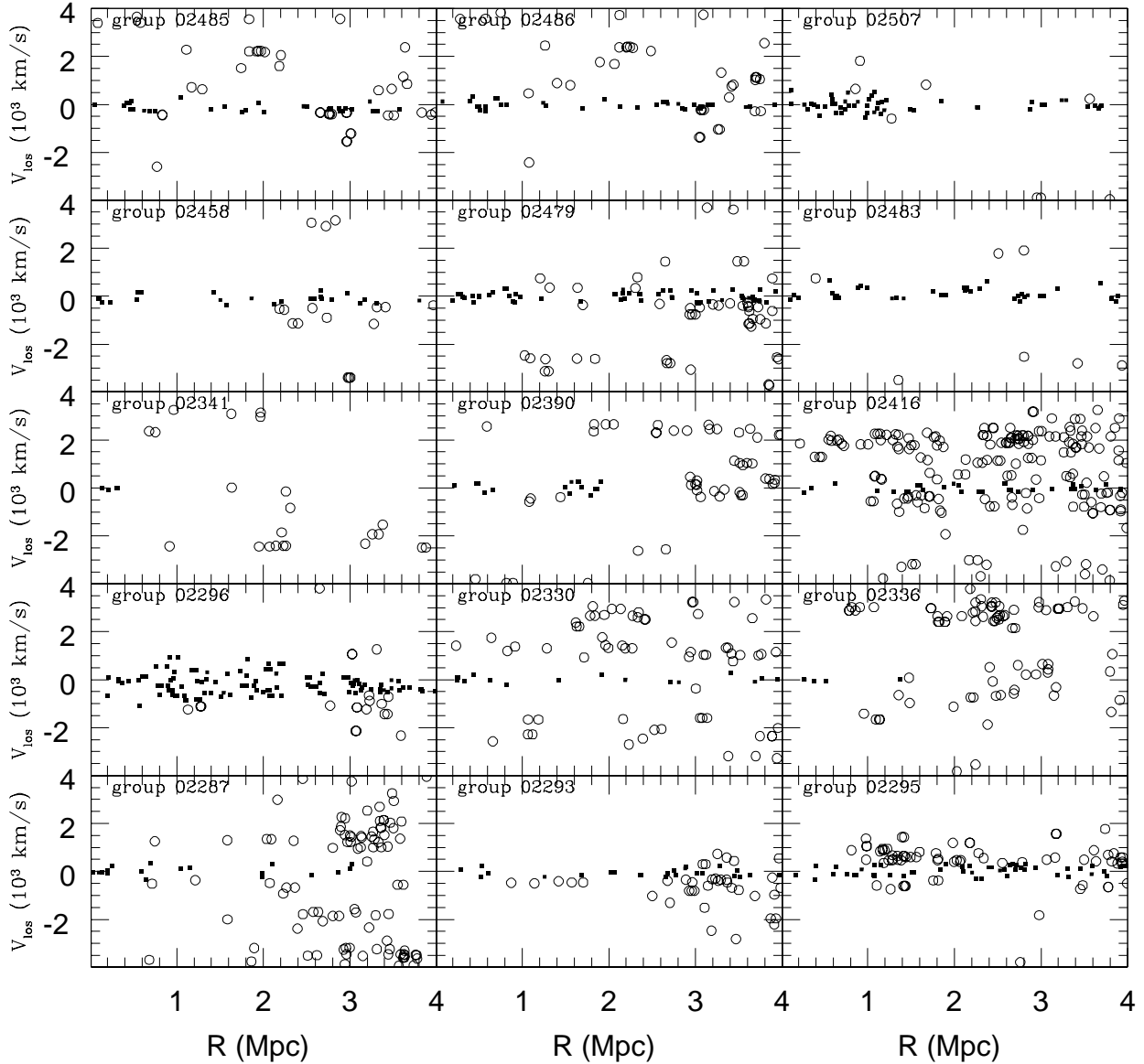}
\end{center}
\caption{Phase-space diagrams of 15  FoF groups shown as examples. The
  velocity, $V_{los}$,  and radial, $R$,  offsets are with  respect to
  the  group centre.   We apply  a shifting  gapper procedure  for the
  selection  of  group  members  (filled  squares)  and  exclusion  of
  interlopers (open circles). Notice  that the plots show all galaxies
  (not only the ETGs) selected from SDSS around each group.  }
\label{fig:shift_gapper}
\end{figure*}

In Fig.~\ref{fig:shift_gapper}  we show  the output of  the ``shifting
gapper''  technique.  Phase-space  diagrams are  exhibited for  15 FoF
groups as an example. The velocity and radial offsets are with respect
to the group centre. Group members are shown with solid squares, while
interlopers  are  plotted  with   open  circles.   This  figure  helps
illustrating the fact that most  systems in the FoF group catalogue have
low velocity dispersions.

\subsection{Selection Effects: limiting magnitude, maximum radius, borders 
of the survey and fiber collision}

To  measure properties  of  the  galaxies and  the  groups where  they
reside, such  as velocity dispersion,  physical radius and  mass, care
should  be taken due  to a  few selection  effects. For  instance, the
limiting magnitude  of the survey  and the maximum  radius, $R_{max}$,
from which  galaxies are selected play a  crucial role. \citet{lop09a}
show  that  a survey  complete  to  $M^*+1$  is sufficient  to  derive
unbiased  estimates  of   velocity  dispersion,  physical  radius  and
mass. They  show that  missing galaxies  in the range  $M^*+1 <  M_r <
M^*+2$  have  no significant  impact  on  cluster  parameters. On  the
contrary, missing galaxies at $M^* < M_r < M^*+1$ affect the estimates
of these  parameters. For the current  work, we only  use galaxies and
groups at  $z < 0.1$.   In that regime,  SDSS is complete to  at least
$M^*+1$.

As discussed by \citet{fad96} and \citet{gir96, gir98}, only values of
velocity  dispersion  obtained  from  all cluster  galaxies  within  a
sufficiently   large   radius   can   be  considered   as   physically
meaningful. The choice  of $R_{max}$ may be difficult  as one wants to
cover  the whole cluster,  but avoid  the effects  of the  large scale
structure. This choice  may also be affected by the  area of a pointed
observation or  the limits of a  large scale survey.   On that matter,
\citet{lop09a} show that a maximum radius of 2.5 $h^{-1}$ Mpc (3.3 Mpc
for $h =$  0.75) is large enough to avoid  biased estimates of cluster
properties. Smaller radius may  lead to overestimation of the velocity
dispersion, while larger radius  may imply a contamination from nearby
structures. For  large values of $R_{max}$, some  assumptions done for
the virial analysis (see below) may  not hold. It is also important to
stress  that the  magnitude and  radial  limits will  also affect  the
membership  selection, and  hence  the galaxy  density estimates  (see
below).

The distance from  a group to the  edge of the survey also  has a role
when estimating group properties.  A very large value of $R_{max}$ has
a  large chance  of overlapping  with  a border.   However, under  the
assumption  of circular  symmetry, partially  missing galaxies  in one
side of  a group do  not affect the  process of selecting  members and
estimating the velocity dispersion.  For this work, we set the maximum
radius  for   membership  selection  to  $R_{{\rm   max}}=$  4.0  Mpc,
consistent with  \citet{lop09a}.  Note  that this choice  for $R_{{\rm
    max}}$ and  the use of  clusters at $z  < 0.1$, result  in cluster
properties   in   good   agreement   to  previous   results   in   the
literature~\citep{lop09a}. \citet{lop09b} also showed that the scaling
relations derived from these parameters provide excellent mass tracers
and are in agreement to previously reported results.

One further issue  affects the SDSS spectroscopic sample  and may have
an impact on  group parameters, as well as galaxy  densities. Due to a
mechanical restriction,  spectroscopic fibers cannot  be placed closer
than 55 arcsecs on the sky.  In case of a conflict, the algorithm used
for  target  selection randomly  chooses  which  galaxy  gets a  fiber
(Strauss et al. 2002). This  problem is reduced by spectroscopic plate
overlaps,   but   fiber  collisions   still   lead   to  a   $\sim$6\%
incompleteness in  the main galaxy  sample.  Obviously, this  issue is
more important  in high galaxy density  regions, such as  the cores of
groups and  clusters. Our  approach to  fix it is  similar to  the one
adopted by Berlind  et al. (2006). For galaxies with $r  <$ 18 with no
redshifts we  assume the redshift of  the nearest neighbour  on the sky
(most of times  the galaxy it collided with). This  may result in some
nearby galaxies to be placed at high redshift, artificially increasing
their  estimated  luminosities.   Hence,  the collided  galaxies  also
assume the magnitudes (in addition  to the redshifts) of their nearest
neighbours,  resulting in  an unbiased  luminosity  distribution.  {
  Notice that this procedure to  take the fiber collision effect into account might be further
  improved  by  taking advantage  of  photo-z  and surface  brightness
  information  of  collided  galaxies.   However, we  decided  not  to
  include this extra information in the correction since (i) the fraction
  of  fixed  galaxies  is  quite  small (at  most  $6\%$,  at  highest
  densities); (ii) the above correction procedure is the same used to
  create the FoF group catalogue, and it has shown to accurately match
  the     multiplicity     function      of     groups     in   the  mock
  catalogues~\citep{Berlind:06}.  In the  end, we measure densities by
  fixing the collision effect in a consistent  way to that used for defining the
  group catalogue.}

\begin{figure}
\begin{center}
\leavevmode
\includegraphics[width=3.5in]{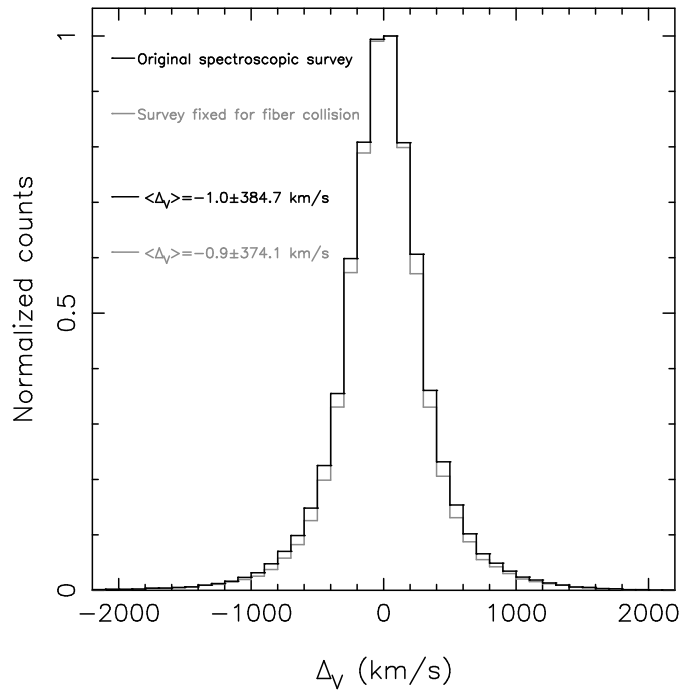}
\end{center}
\caption{Velocity  offset  (relative  to  group  centre)  distribution
  before and after fixing the fiber collision issue.}
\label{fig:vel_off}
\end{figure}

To show that  the velocity distribution is not affected  by the way we
apply this  correction we show in  Fig.~\ref{fig:vel_off} the velocity
offset distribution (relative to group centre) before and after fixing
the fiber collision issue.  The  mean values of the two distributions,
with the corresponding uncertainties, are indicated in the bottom left
of the figure.

\section{Group Properties}
\label{sec:group_properties}
In  this  section  we  briefly  describe the  procedures  adopted  for
estimating  group properties,  such as  velocity  dispersion, physical
radius  ($R_{200}$),  mass  ($M_{200}$),  and  substructure.   Further
details  can  be  found   in  \citet{lop09a}.   These  parameters  are
considered when studying the dependence of the FP on the environment.

\subsection{Virial Analysis}

For the  virial analysis, first we compute  the line-of-sight velocity
dispersion  ($\sigma_{P}$) of  all group  members within  the aperture
R$_A$ (the radial  offset of the most distant  group member). R$_A$ is
normally close to  the maximum radius ($R_{{\rm max}}=$  4.0 Mpc), but
could be  much smaller in case  the outermost galaxies  are flagged as
interlopers. The robust velocity dispersion estimate ($\sigma_{P}$) is
given by the  gapper or bi-weight estimator, depending  on whether the
number  of  galaxies  available  is   $<  15$  (gapper)  or  $\ge  15$
(bi-weight; \citealt {bee90}).   Following the prescriptions of \citet
{dan80}, the velocity dispersion is corrected for velocity errors.  We
also obtain  an estimate of  the projected ``virial  radius'' R$_{PV}$
\citep  {gir98}. A  first  estimate of  the  virial mass  is given  by
(equation 5 of \citealt {gir98})

\bea
{\rm M_V} = {3\pi \sigma^2_{P} R_{PV} \over
             2 G},\,\,\,
\label{eq:perpdef}
\eea  where  G is  the  gravitational  constant  and $3\pi/2$  is  the
de-projection factor.

Next  we  apply the  surface  pressure  term  correction to  the  mass
estimate \citep {the86}.   To this effect, we need  an estimate of the
R$_{200}$  radius,  which  (as  a  first  guess)  is  taken  from  the
definition of  Carlberg et  al. (1997; equation  8 of that  paper). We
assume  the percentage  errors for  the  mass estimates  are the  same
before  and after  the  surface pressure  correction  (as in  \citealt
{gir98}).

After  applying  this correction,  we  obtain  a  refined estimate  of
R$_{200}$ considering the virial mass density.  If $M_V$ is the virial
mass (after  the surface  pressure correction) in  a volume  of radius
$R_{A}$,    R$_{200}$    is     then    defined    as    R$_{200}    =
R_{A}[\rho_{V}/(200\rho_c(z))]^{1/2.4}$. In  this expression $\rho_{V}
=  3M_V/(4\pi R_{A}^3)$  and $\rho_c(z)$  is the  critical  density at
redshift $z$. The  exponent in this equation is  the one describing an
NFW  \citep{nar97} profile near  R$_{200}$ \citep  {kat04}. If  we use
five-hundred  instead of two-hundred  on this  equation, we  obtain an
estimate  of R$_{500}$.   Next, assuming  an NFW  profile  we estimate
M$_{200}$  (or  M$_{500}$)  from  the interpolation  (most  cases)  or
extrapolation of the  virial mass M$_V$ from R$_{A}$  to R$_{200}$ (or
R$_{500}$). Then,  we use the  definition of M$_{200}$  (or M$_{500}$)
and  a final  estimate of  R$_{200}$ (or  R$_{500}$) is  derived. This
procedure is  analogous to what is  done by \citet  {pop07} and \citet
{biv06}.

\begin{figure*}
\begin{center}
\leavevmode
\includegraphics[width=7.2in]{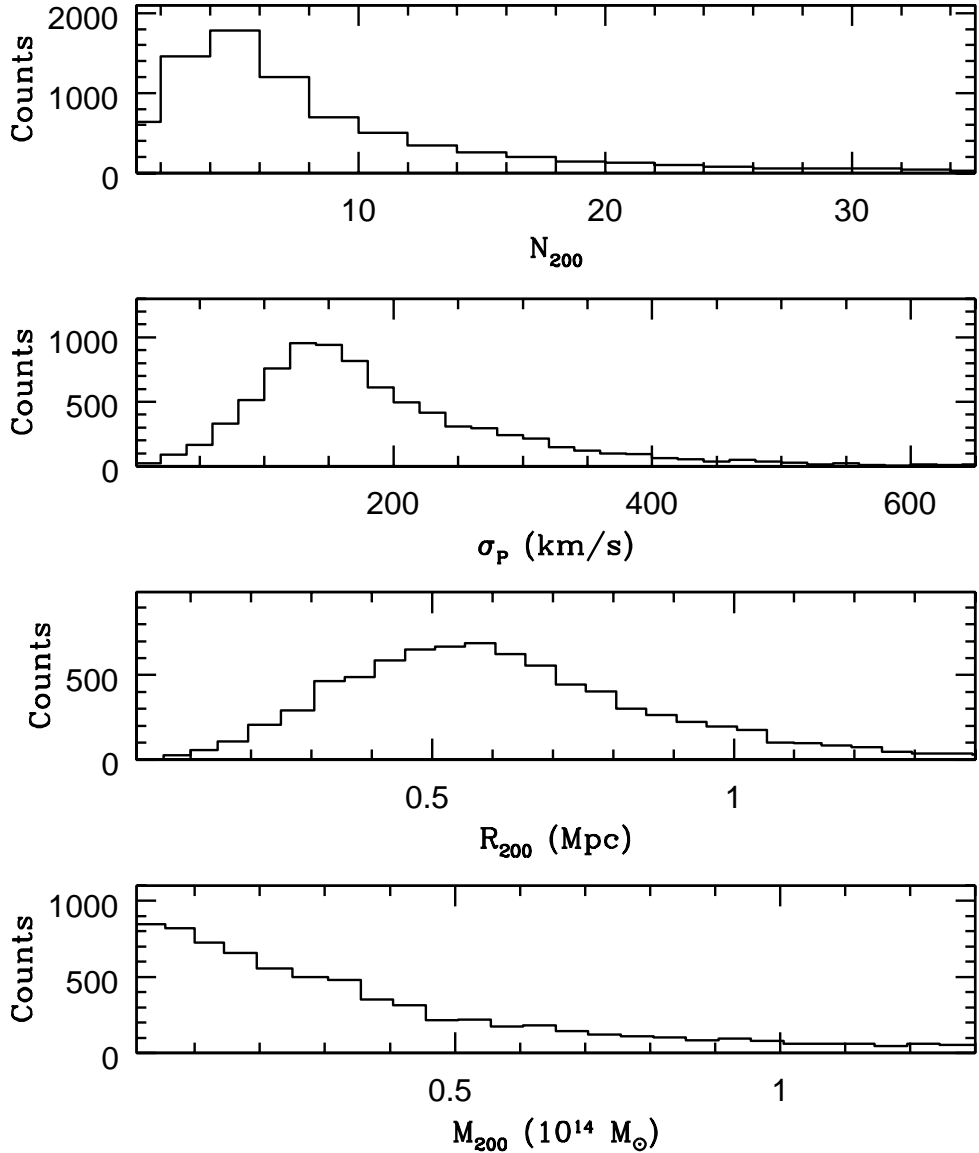}
\end{center}
\caption{Distributions of N$_{200}$  (number of member galaxies within
  R$_{200}$), velocity  dispersion (in km/s), R$_{200}$  (in Mpc), and
  M$_{200}$ ($10^{14}$  M$_{\odot}$). Notice that most  of systems are
  poor groups, though there is  a conspicuous number (101) of clusters
  with $\sigma_P$ higher than $600 \, km s^{-1}$.}
\label{fig:dist_group_prop}
\end{figure*}

Fig.~\ref{fig:dist_group_prop}   shows,  from   top  to   bottom,  the
following distributions:  N$_{200}$ (number of  member galaxies within
R$_{200}$);  velocity dispersion  (in km/s);  R$_{200}$ (in  Mpc); and
M$_{200}$ ($10^{14}$ M$_{\odot}$). Note that, as expected from the FoF
selection  function,  most  systems  are  low richness  and  low  mass
systems.  However, there are  still 101  (112) clusters  with velocity
dispersion (mass)  larger than $600  \, km s^{-1}$ ($5  \times 10^{14}
M_\odot$).  Six clusters  have $\sigma_P > 1000 km/s$,  with a maximum
value of $1591 \, km s^{-1}$.

{  In a  forthcoming contribution  of  the SPIDER  series, we  will
  analyse  the scaling  relations of  groups and  clusters in  the FoF
  catalogue (cf.~\citealt{lop09b})}.

\subsection {Substructure estimates}
\label{sec:substructures}

On what regards substructure, we  employ a 3D substructure test to all
FoF groups with at least five galaxy members within R$_{200}$. The DS,
or $\Delta$  test \citep{dre88}, is applied as  in \citet{lop09a}. The
code  used  for the  substructure  analysis  is  the one  from  \citet
{pin96},  who  evaluated  the  performance of  thirty-one  statistical
tests.

The  algorithm  computes  the  mean velocity  and  standard  deviation
($\sigma$) of  each galaxy and  its N$_{nn}$ nearest  neighbours, where
N$_{nn} =  $ N$^{1/2}$ and  N is the  number of galaxies in  the group
region. Then  these local mean  and $\sigma$ values are  compared with
the  global mean  and  $\sigma$  (based on  all  galaxies). For  every
galaxy, a  deviation from the global  value is defined  by the formula
below.   Substructure  is  estimated  with  the  cumulative  deviation
$\Delta$ (defined by $\sum \delta_i$;  see below). For objects with no
substructure we have $\Delta \sim$ N.

\bea
{\rm \delta_i^2} = \left( \frac{N_{nn}+1}{\sigma^2} \right)[(\bar{v}_{local}-\bar{v})^2 + (\sigma_{local} - \sigma)^2]. \,\,\,
\eea

Any  substructure test statistic  has little  meaning if  not properly
normalised, which  can be  achieved by comparing  the results  for the
input  data to  those  for substructure-free  samples  (the {\it  null
  hypothesis}).  For  the $\Delta$ test the null  hypothesis files are
created after the velocities are shuffled randomly with respect to the
positions, which remain fixed \citep{pin96}.

For  each   input  data   set,  we  generate   five-hundred  simulated
realisations.  We then calculate the number of Monte Carlo simulations
which show more substructure than  the real data. Finally, this number
is  divided by  the  number of  Monte  Carlo simulations.  We set  our
significance  threshold  at   5$\%$,  meaning  that  only  twenty-five
simulated data  sets can have substructure statistics  higher than the
observations to consider  a substructure estimate significant. Further
details   can  be   found   in  \citet   {pin96},  \citet{lop06}   and
\citet{lop09a}.

For  each  FoF  group the  substructure  test  is  applied to  the  3D
distribution of all galaxy members, as selected by the shifting-gapper
method. The  $\Delta$ test  is only applied  to systems with  at least
five  galaxies available  within R$_{200}$.  Out of  the  8,083 groups
satisfying  these criteria, we  find that  3,570 ($\sim$  44,2\%) show
significant  signs of  substructure.  We note  that  this fraction  is
larger than found by \citet{lop09a}.  That is probably due to the fact
that  (in  order  to  correct  for  the  fiber  collision  issue)  we
artificially   assign   redshifts  to   galaxies   not  targeted   for
spectroscopic  observations. 

{ We found that the fraction of groups with substructures increases from
$\sim  30\%$ at $M_{200}  \sim 10^13  M_{\odot}$ to  $\sim 80  \%$ for
clusters as  massive as $10^{14}  M_{\odot}$. Notice that  this result
should be taken  with caution, as low (relative  to high) mass systems
usually have a  lower number of member galaxies  and this might affect
the   efficiency  in   detecting   substructures  at   the  low   mass
regime~\footnote{This  issue  will  be  addressed in  the  forthcoming
  contribution where  we analyse the  scaling relations of  groups and
  clusters in the FoF catalogue.}}.

\section{The Environment of ETGs}
\label{sec:environment}
\subsection {Density Estimates}

Galaxy density  is estimated as follows.  For each galaxy  member of a
given FoF group, we compute  its projected distance, d$_N$, to the Nth
galaxy of  that group, where  N = N$_{memb}^{1/2}$, and  N$_{memb}$ is
the  number  of member  galaxies  in  the  group.  The  local  density
$\Sigma_N$  is  given  by   N/$\pi$d$_N^{2}$,  measured  in  units  of
galaxies/Mpc$^2$.  The d$_N$ is expressed in units of $Mpc$, using the
redshift  of the  given group.   For  the density  estimation we  only
consider member galaxies within a fixed luminosity range (see comments
below).   Notice that  differently  from other  works  density is  not
computed  relative  to all  other  galaxies  in  the field,  but  only
relative  to  the  other  group  members.   Hence,  instead  of  using
$\Sigma_5$ or $\Sigma_{10}$, which  are more common in the literature,
we use  $\Sigma_N$ so  that the distance  used for  estimating density
scales according to the number  of members (or group mass).  A further
note  is that if  for a  given galaxy  there are  fewer than  N member
neighbours  (in the  given  luminosity range),  we  cannot compute  the
distance  to the  Nth  galaxy of  that  group, d$_N$.   In that  case,
$\Sigma_N$  is set to  a lower  limit of  $0.01$.  However,  that only
happens for about $0.1\%$ of the objects in the optical sample of ETGs
with group membership.

In \citet{LBdC10b} (hereafter paper II),  we showed that the sample of
ETGs is complete to $M_r  = -20.55$ for 2DPHOT total magnitudes.  This
corresponds to  a limit of $M_r  = -20.32$ for  SDSS model magnitudes.
Here  we compute local  density by  (SDSS) petrosian  magnitudes.  The
above  limit translates to  $M_r =  -20.23$ for  petrosian magnitudes.
The density estimates are derived in the same luminosity range for all
galaxies,  no matter which  FoF group  they belong  to. This  limit is
fixed  by the  maximum  redshift of  SPIDER  ($z =  0.095$).  At  this
redshift,  for petrosian magnitudes,  the characteristic  magnitude of
the galaxy luminosity function is $M^* = -21.43$ in r band (Popesso et
al.  2005).  If we make $M^* +  X = -20.23$, we find $X = 1.2$. Hence,
when deriving  densities, we select  only group members  brighter than
$M^* + 1.2$.

\begin{figure}
\begin{center}
\leavevmode
\includegraphics[width=3.5in]{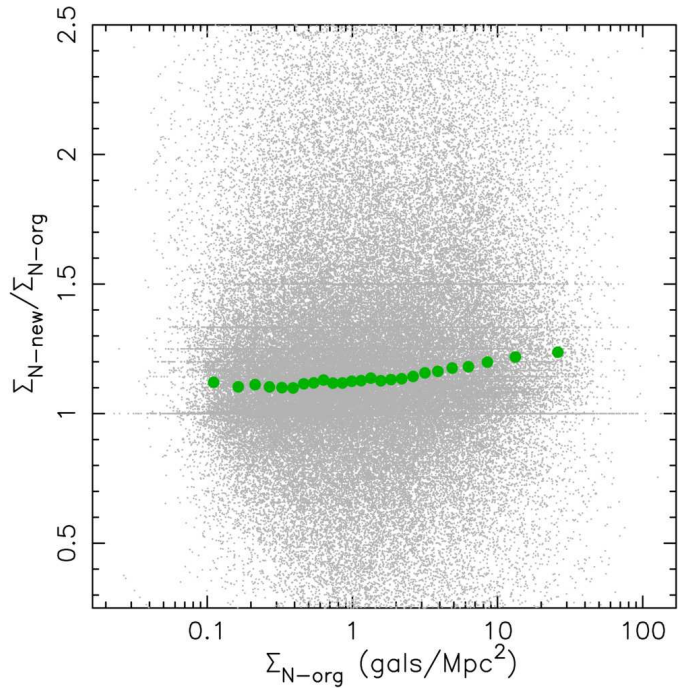}
\end{center}
\caption{Comparison   of  $\Sigma_N$   computed   with  the   original
  spectroscopic  SDSS  survey  and  after  fixing it  for  the  fiber
  collision  issue. Dots  are  used  for all  the  galaxies for  which
  density is estimated. Large green points show the median values for
  bins containing 5000 galaxies each.}
\label{fig:comp_sigma}
\end{figure}

\begin{figure}
\begin{center}
\leavevmode
\includegraphics[width=3.5in]{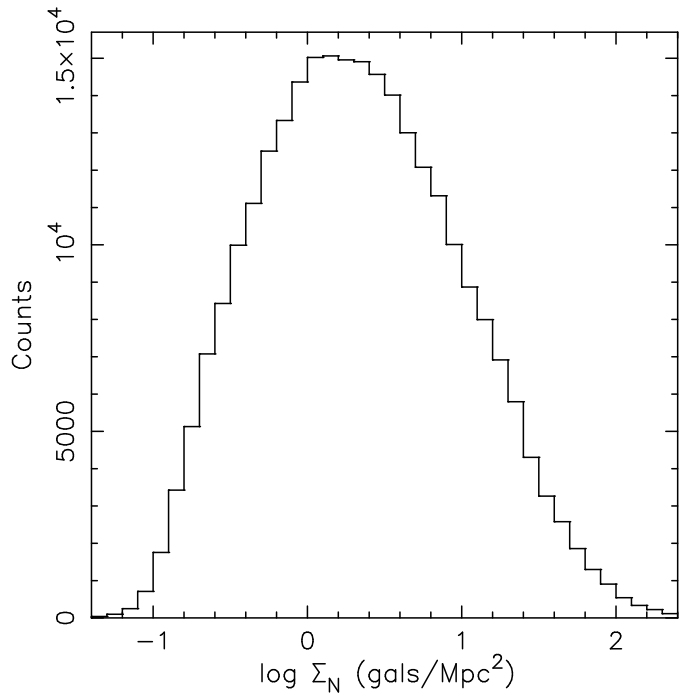}
\end{center}
\caption{Distribution  of local  galaxy density  ($\Sigma_N$)  for all
  group members.}
\label{fig:dens_dist}
\end{figure}

Fig.~\ref{fig:comp_sigma} shows the  comparison of $\Sigma_N$ computed
with the  original spectroscopic SDSS  survey and after fixing  it for
the fiber  collision issue (see section  2.2). In this  Figure, we use
dots  to show  the  results for  all  galaxies for  which density  was
estimated.   Large  green  points  show  the median  values  for  bins
containing  $5,000$ galaxies.  As  expected, the  correction increases
the  $\Sigma_N$  value.   The  effect  is slightly  stronger  at  high
$\Sigma_N$, the regime where we  expected the fiber collision issue to
be   more  relevant.    Fig.~\ref{fig:dens_dist}  shows   the  density
distribution for  all member galaxies  within the $8,083$  FoF groups.
Galaxies residing in the FoF groups span a wide range of environments,
corresponding to more than three decades in local galaxy density.

\subsection {Definition of group and field galaxies}
\label{sec:field_group_samples}

The environment of  ETGs is first defined by the  location of a galaxy
in either a  group or the field. For galaxies  residing in groups, the
environment is further characterized in terms of the group properties,
such  as  the  group-centric  distance,  mass,  physical  radius,  and
substructure.   Group  galaxies  have  also  local  density  estimated
relative to other group members  (as described above), while for field
galaxies,  we use  the FP  relation itself  to introduce  a fictitious
local density value (see Sec.~\ref{sec:offset_rband}).

Galaxies residing  in groups are simply those  having group membership
according to  the shifting gapper method (within  $R_{{\rm max}}=$ 4.0
Mpc;   see  Sec.~\ref{sec:group_catalog}).    Both  the   optical  and
optical+NIR    samples    of    ETGs   (Sec.~\ref{sec:samples})    are
cross-correlated to the list of  galaxies belonging to groups, so that
we  define which  SPIDER ETG  has a  parent group  in the  FoF updated
catalogue.

Field ETGs are  defined as follows.  We start  selecting only the ETGs
that are  not members  of any  group in the  updated FoF  catalogue.  As
reported in Sec.~\ref{sec:group_catalog}, because of the low number of
galaxies  available for  some (poor)  groups, we  are able  to measure
group properties  and define  galaxy members for  $79.2\%$ of  the FoF
groups. This  implies that some  galaxies with no  assigned membership
may  still reside in  one of  the not-measured  groups, and  should be
excluded from a genuine field  sample.  To clean the field sample from
these cases,  we check all  the FoF groups (10,124  systems), removing
those ETGs with radial offsets  $< 5 \times R_{perp,rms}$ and velocity
offsets $c \times \Delta_z <  5 \times \sigma_v$ from any group, where
$R_{perp,rms}$  and $\sigma_v$  are the  projected rms  radius  of the
group  and  velocity  dispersion   as  defined  in  the  original  FoF
catalogue~\citep{Berlind:06}.  Galaxies on the  border of a group in the
phase-space   diagrams   (Fig.~\ref{fig:shift_gapper})   can  be   not
classified  as group  members by  the shifting  gapper  technique, but
nevertheless  their properties might  still be  affected by  the group
because  of the  in-falling in  its  potential well.   To remove  this
further source of $``contamination''$  from the field sample, we check
once more all the 8,083  measured groups, removing objects with radial
offsets $< 5 \times R_{200}$ and velocity offsets $c \times \Delta_z <
5 \times \sigma_P$ from any  group, where $R_{200}$ and $\sigma_P$ are
the characteristic radius and  velocity dispersion of a group computed
as described  in Sec.~\ref{sec:group_properties}.  The  remaining ETGs
define the field sample.

Applying this procedure to the  r-band sample of ETGs ($N=36,124$), we
obtain   two  subsamples   of  $16,717$   group  and   $11,824$  field
galaxies. For  the optical+NIR sample  of ETGs ($N=4,589$),  the above
selection  results  in  $2,187$  group  and  $1,359$  field  galaxies,
respectively.

\section{Measuring the FP relation}
\label{sec:fp}
As in paper II, we write the FP relation as:
\begin{equation}
\log R_{\rm e} = a \log \sigma_{\rm 0} + b < \! \mu \! >_{\rm e} + c,
\label{eq:FP}
\end{equation}
where $''a''$  and $''b''$ are the  slopes, and $''c''$  is the offset
(intercept) of the FP, \lre  \, is the logarithmic effective radius of
galaxies,  in units of  kpc, \mie  \, is  the mean  surface brightness
within   that  radius,   and  $\sigma_0$   is  the   central  velocity
dispersion. For the present study, we use $\sigma_0$ values from SDSS,
although we  verified that results  are the same when  using STARLIGHT
velocity dispersions (see paper  I). The $\sigma_0$'s are corrected to
a  circularised aperture  of radius  $r_e/8$, following~\citet{JFK95}.
The computation of  \lre \, and \mie \, in  the different wavebands is
described  in detail  in paper  I. In  general, we  refer  to $''a''$,
$''b''$, and $''c''$, as the coefficients of the FP.

In  order to  analyse  environmental effects  on  FP coefficients,  we
proceed as follows.  In Sec.~\ref{sec:offset_fp}, we first compare the
offset  of the  FP in  different  environmental bins.   The values  of
$``c''$  for samples of  ETGs in  the different  bins are  computed by
fixing the  slopes of the FP  to the values of  $a=1.39$ and $b=0.315$
obtained in paper II for the  entire SPIDER sample.  This allows us to
interpret  differences  in  $``c''$  as  differences  in  the  average
mass-to-light  (hereafter   $M/L$)  ratio  of   ETGs  among  different
environmental bins  (Sec.~\ref{sec:offset_sp}). { Notice  that this
  FP-based $M/L$  ratio differs  from the quantity  $M_{dyn}/L$, where
  dynamical   mass~\footnote{    In   order   to    determine   the
    proportionality  factor relating  $M_{dyn}$ and  $R_e \sigma_0^2$,
    one has  to assume a  given galaxy model, making  some assumptions
    about the  (three-dimensional) distribution of  dark- and luminous
    matter in a galaxy. On the other hand, the FP-based $M/L$ does not
    have to rely  on any of these assumptions.},  $M_{dyn} \propto R_e
  \sigma_0^2$, and galaxy luminosity, $L$, are directly estimated from
  the data (see e.g.~\citealt{Bern:03a}). } The $``c''$ is measured as
\begin{equation}
 c =  \overline{\log R_e} - a \times \overline{\log \sigma_0} -b \times \overline{<\mu>_e},
\label{eq:c}
\end{equation}
where \olre, \omie,  and \ols \, are the median  values of \lre, \mie,
and \ls.  { For a given  sample of ETGs (i.e. a given environmental
  bin), the uncertainties on $''c''$  are computed by the width of the
  distribution of $''c''$ values,  as computed by Eq.~\ref{eq:c}, from
  $1000$ bootstrap  realisations of the  sample.  Since the  values of
  $''a''$  and  $''b''$  are  kept  fixed in  Eq.~\ref{eq:c}  for  all
  realisations, this procedure neglects the effect of the (correlated)
  uncertainties   on   $''a''$   and   $''b''$.    As   discussed   in
  Sec.~\ref{sec:offset_fp}, this further  source of uncertainty can be
  completely  neglected, as  far as  the difference  of  $''c''$ among
  different environments  is concerned.}  The  environmental variation
of $``c''$ is used in Sec.~\ref{sec:slopes_fp} to properly measure the
slopes  of the  FP  for  the ETG's  samples  populating the  different
(environmental) bins.  The values of $``a''$ and $``b''$ are estimated
by an orthogonal robust  fitting procedure and corrected for selection
effects  as  well  as   the  effect  of  correlated  uncertainties  on
structural  parameters (see  paper II  for details).   We  implement a
procedure  to  account  for  the different  average  mass-to-light  of
galaxies  in  the  different   environmental  bins,  as  well  as  the
$``$luminosity   segregation$''$  of   galaxy  populations   with  the
environment (see  Sec.~\ref{sec:slopes_fp} for details).   The scatter
of  ETGs  around the  FP  and its  dependence  on  environment is  not
analysed  here, being  the  subject  of a  forthcoming  paper in  this
series.

\section{The offset of the FP as a function of environment}
\label{sec:offset_fp}
  We start analysing the dependence of the offset $``c''$ of the FP on
  galaxy environment. Sec.~\ref{sec:offset_rband} presents the results
  obtained from  the optical  sample of ETGs,  while the  variation of
  ``c''     from     $g$     through     $K$    is     analysed     in
  Sec.~\ref{sec:offset_optNIR}.

\subsection{The FP offset in r band}
\label{sec:offset_rband}

\begin{figure}
\begin{center}
\includegraphics[height=80mm]{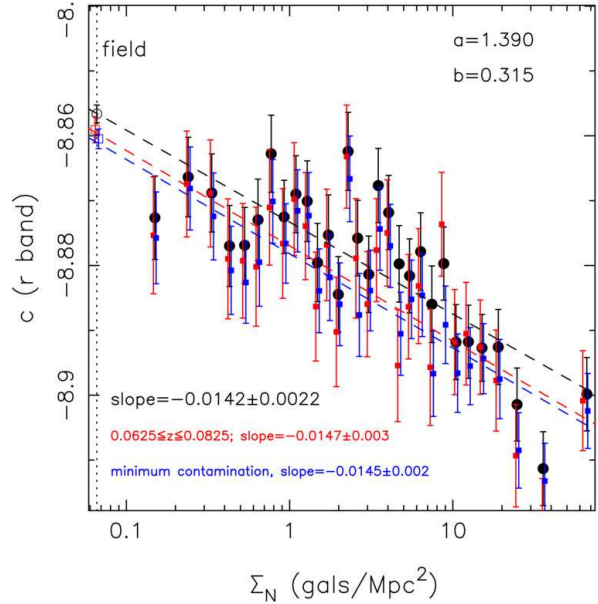}
\caption{Variation of  the offset $``c''$ of  the FP as  a function of
  local galaxy density, $\Sigma_N$.  Filled circles plot the values of
  $``c''$ obtained  binning the group  sample of ETGs with  respect to
  $\Sigma_N$, adopting 30 bins with  the same number of galaxies each.
  Error bars  denote $1~\sigma$  uncertainties.  The $``c''$  has been
  computed  by fixing  $``a''$ and  $``b''$ in  all bins  to  the same
  values  of  $1.39$  and  $0.315$,  respectively,  as  shown  in  the
  upper-right  corner of  the figure.   The dashed  black line  is the
  best-fit  line to  the filled  circles, derived  by  a least-squares
  fitting  procedure with  $``c''$  as dependent  variable. The  slope
  value of  the line is reported  in the lower--left of  the plot. The
  value of $``c''$  for the field sample of ETGs is  also shown in the
  figure as an empty circle.  The average local density value of field
  galaxies, marked by  the vertical dotted line, is  assigned by using
  the  above linear  fit (see  the text).   Red and  blue  squares are
  obtained by selecting only galaxies in a narrower redshift range and
  in  the  lower  contamination  sample of  ETGs.   The  corresponding
  least-squares fits of $``c''$  versus $\log \Sigma_N$ are plotted by
  the red  and blue dashed lines,  whose slope values  are reported in
  the lower--left of the Figure.
\label{fig:offset_r_dens}
}
\end{center}
\end{figure} 

Fig.~\ref{fig:offset_r_dens}  plots  the  value  of  $``c''$  for  the
optical  sample of ETGs,  as a  function of  the local  galaxy density
$\Sigma_N$. The  sample of  group galaxies is  binned with  respect to
$\Sigma_N$, with each bin including the same number of $557$ galaxies.
This  leads to  $30$  density  bins.  The  Figure  shows that  $``c''$
smoothly decreases from low to  high density regions.  A linear fit of
this trend, with
\begin{equation}
c = a_{\Sigma_N} + b_{\Sigma_N} \log \Sigma_N,  
\label{eq:c_dens}
\end{equation}
gives $a_{\Sigma_N}= -8.8734 \pm 0.0018$ and $b_{\Sigma_N}= -0.014 \pm
0.002$.   The  value of  $b_{\Sigma_N}$  differs  from  zero at  $\sim
7$~$\sigma$, meaning that the  variation of $``c''$ with $\Sigma_N$ is
highly  significant.   The   quantity  $b_{\Sigma_N}$  represents  the
variation of $``c''$  per decade of local galaxy  density, and encodes
all  the information  about the  dependence of  the FP  offset  on the
(local)  environment  (see  Sec.~\ref{sec:offset_sp}). For  the  field
sample of  ETGs, an  estimate of the  local density is  not available.
Hence, we  assign a fictitious $\Sigma_N$  value, $\Sigma_{field}$, to
all the field ETGs.  To this aim, we first compute the $``c''$ for the
field    sample.     Then,    inverting   Eq.~\ref{eq:c_dens},    with
$a_{\Sigma_N}$ and $b_{\Sigma_N}$ given by the values listed above, we
infer the $\Sigma_{field}$ value.

\begin{table}
\centering
\small
\begin{minipage}{70mm}
 \caption{Covariance matrix terms of the uncertainties on $a$, $b$,
$a_{\Sigma_N}$, and $b_{\Sigma_N}$.}
  \begin{tabular}{c|c|r}
   \hline
    $X$ & $Y$ &  $C(X,Y)$ \\
   \hline
 $a           $ & $a           $ & $ 4.150e-04$ \\
 $a           $ & $b           $ & $ 9.964e-06$ \\
 $a           $ & $a_{\Sigma_N}$ & $-1.515e-03$ \\
 $a           $ & $b_{\Sigma_N}$ & $-2.900e-06$ \\
 $b           $ & $b           $ & $ 1.970e-06$ \\
 $b           $ & $a_{\Sigma_N}$ & $-7.462e-05$ \\
 $b           $ & $b_{\Sigma_N}$ & $-3.499e-08$ \\
 $a_{\Sigma_N}$ & $a_{\Sigma_N}$ & $ 6.378e-03$ \\
 $a_{\Sigma_N}$ & $b_{\Sigma_N}$ & $ 7.565e-06$ \\
 $b_{\Sigma_N}$ & $b_{\Sigma_N}$ & $ 5.137e-06$ \\
   \hline
  \end{tabular}
\label{tab:cov_mat_abcd}
\end{minipage}
\end{table}

A possible source of concern  in the trend of $``c''$ with environment
is  the  contamination  of  the  SPIDER  samples  from  galaxies  with
non-genuine  early-type   morphology  (see  paper  I).    Due  to  the
morphology-density  relation~\citep{Dressler:80},  these  objects  are
expected to  be more abundant in  the low density  regions. To analyse
the impact  of such contamination,  in paper I  we have defined  a low
contamination sample of ETGs  consisting of $32,650$ (out of $39,993$)
galaxies. For  each bin of local  density, we select  only galaxies in
this low contamination subsample.  The resulting trend of $``c''$ with
local density is  shown by the blue symbols in  the Figure.  Besides a
rigid shift, the slope of the trend is identical to that obtained with
the entire sample,  proving that contamination does not  affect at all
our results.  Another  source of concern is the  redshift range of the
SPIDER sample.   Fainter galaxies are relatively more  abundant in the
lower density regions, and are  better represented in the low redshift
tail  of  the  sample.   Due  to small  evolutionary  effects  in  the
luminosity  of ETGs (e.g.~\citealt{Bern:03a}),  this effects  can bias
the trend  of the  offset with the  environment.  To tackle  with this
issue,  we select in  all the  environmental bins  only galaxies  in a
narrower   redshift   bin,  from   $z=0.0625$   to  $z=0.0825$.    The
corresponding trend of $``c''$ with  local density is shown by the red
symbols   in   Fig.~\ref{fig:offset_r_dens}.    The   slope   of   the
$``c''$--$\log  \Sigma_N$ fit  is fully  consistent with  that  of the
entire sample.

We find that the $``c''$--$\log \Sigma_N$ trend is robust with respect
to  the statistical  estimator  used to  calculate  the FP  intercept.
Replacing the median values of \lre, \mie\, and \ls, in Eq.~\ref{eq:c}
with  the  peak values  of  the  distributions,  as estimated  by  the
bi-weight  estimator~\citep{bee90}, we  obtain a  very  similar trend,
whose slope, $b_{\Sigma_{N}}$, is indistinguishable from that reported
in Fig.~\ref{fig:offset_r_dens}.  Since the bi-weight (relative to the
median) estimator is  less sensitive to the shape of  the tails in the
distribution being  analysed, this test  also shows that the  trend of
$``c''$ with $\log \Sigma_N$ is not affected by the different shape of
the  massive  end  of  the  galaxy luminosity  function  in  different
environments, i.e.  the larger fraction of  bright galaxies inhabiting
higher  (relative to  lower)  density environments.   Notice that  the
values of $``c''$ in different bins of $\log \Sigma_N$ are measured by
keeping  the   values  of  FP  slopes,  $``a''$   and  $``b''$,  fixed
(Sec.~\ref{sec:fp}).     On   the    other   hand,    as    shown   in
Sec.~\ref{sec:slopes_fp}, the FP slopes  show a small, but significant
trend  with  the  environment,  which might  bias  the  $``c''$--$\log
\Sigma_N$  trend. In  order to  address the  issue, we  recomputed the
$``c''$  by  setting  $a=2$   and  $b=0.4$  in  Eq.~\ref{eq:FP}.  This
corresponds to measure the mass-to-light ratio of ETGs by means of the
virial  theorem,  rather than  the  FP  equation,  hence avoiding  the
possible  $``degeneracy''$ between the  environmental variation  of FP
slopes and intercept. Setting $a=2$  and $b=0.4$, the trend of $``c''$
with $\log \Sigma_N$ turns out  to be fully consistent with that shown
in  Fig.~\ref{fig:offset_r_dens}, with the  slope of  the best-fitting
line  amounting  to  $-0.02   \pm  0.004$  (compared  to  $-0.014  \pm
0.002$).  More in  general,  we verified  that  all the  environmental
trends of $``c''$ remain statistically unchanged when using either the
FP or the virial theorem equation.

\begin{figure}
\begin{center}
\includegraphics[height=80mm]{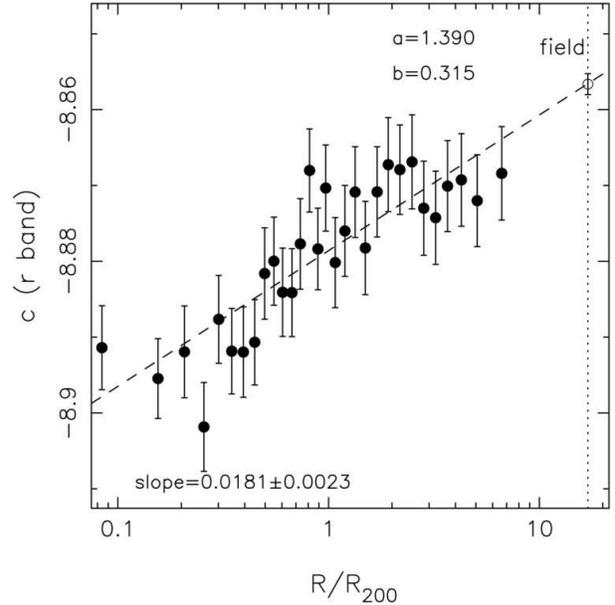}
\caption{Same as Fig.~\ref{fig:offset_r_dens}  but plotting ``c'' as a
  function of the normalised cluster-centric distance $R/R_{200}$. The
  trend of increasing ``c''  with $R/R_{200}$ closely reflects the one
  shown in Fig.~\ref{fig:offset_r_dens}  for the local galaxy density.
  Notice that  the field sample  is also plotted  in the Figure  as an
  empty circle.  The cluster-centric distance, $R_{field}/R_{200}$, is
  assigned  to field  galaxies in  the same  way as  the  local galaxy
  density, $\Sigma_{field}$ (see the text).
\label{fig:offset_r_rad}
}
\end{center}
\end{figure} 

{  The error bars  in Fig.~\ref{fig:offset_r_dens}  deserve further
  comment.    As   stated  in   Sec.~\ref{sec:fp},   we  compute   the
  uncertainties  on $''c''$ by  keeping $''a''$  and $''b''$  fixed in
  Eq.~\ref{eq:c}.    This  corresponds  to   neglect  the   effect  of
  (correlated)  errors of  $''a''$ and  $''b''$ on  the error  bars of
  $''c''$ in Fig.~\ref{fig:offset_r_dens},  and hence on the estimated
  uncertainty on  $b_{\Sigma_N}$. To analyse the effect  of the errors
  on $''a''$  and $''b''$, we  adopt the following approach.   We take
  the covariance  matrix of the uncertainties on  $''a''$ and $''b''$,
  as estimated by the orthogonal fitting procedure (see paper II), and
  perform $1000$ iterations, shifting  each time the values of $''a''$
  and  $''b''$  according  to   their  correlated  errors.   For  each
  iteration, the  values of $''c''$  in all the environmental  bins of
  Fig.\ref{fig:offset_r_rad} are re-computed, together with the values
  of $a_{\Sigma_N}$ and $b_{\Sigma_N}$.  Notice that in a given bin of
  $\Sigma_N$,  the  $''c''$  is  computed by  extracting  a  bootstrap
  realisation of the galaxy sample  in that bin.  We then estimate the
  covariance matrix of the values of $''a''$, $''b''$, $a_{\Sigma_N}$,
  and   $b_{\Sigma_N}$,   among   the   different   iterations.    The
  uncertainties provided  by this  covariance matrix account  for both
  the scatter  around the FP,  as already given  by the error  bars in
  Fig.~\ref{fig:offset_r_rad}, as well as  the correlated errors on FP
  slopes.   All terms of  the covariance  matrix, $C(X,Y)$,  where $X$
  ($Y$) is one of the quantities $''a''$, $''b''$, $a_{\Sigma_N}$, and
  $b_{\Sigma_N}$,  are reported  in  Tab.~\ref{tab:cov_mat_abcd}.  The
  root square of the terms $C(a,a)$ and $C(b,b)$ are the uncertainties
  on $''a''$  and $''b''$  in r-band (cf.   tab.~6 of  paper~II).  The
  1~$\sigma$ uncertainty  on $b_{\Sigma_N}$ amounts  to $\sim 0.0023$,
  being       given       by       the      root       square       of
  $C(b_{\Sigma_N},b_{\Sigma_N})$($=5.137e-06$).   This  uncertainty is
  indistinguishable   from   the  value   of   $0.0022$  reported   in
  Fig.~\ref{fig:offset_r_rad}, where the uncertainty on $b_{\Sigma_N}$
  was  estimated  by  neglecting  the  uncertainties  on  $''a''$  and
  $''b''$.   In other  words, as  far  as the  environmental trend  of
  $''c''$ is concerned (i.e.  the value of $b_{\Sigma_N}$), the effect
  of the  errors on $''a''$  and $''b''$ can be  completely neglected,
  and the dominant source of error  is the rms of residuals around the
  FP.  Notice also that the covariance term between $b_{\Sigma_N}$ and
  $''a''$ is negative, implying that as $''a''$ increases the value of
  $b_{\Sigma_N}$  tends to  become  smaller. The  same  holds for  the
  covariance  term $C(b,b_{\Sigma_N})$.  This  is consistent  with the
  fact  that, when fixing  the values  of $''a''$  and $''b''$  to the
  virial  theorem expectations,  we obtain  a value  of $b_{\Sigma_N}$
  slightly  smaller, but still  statistically consistent  (see above),
  than that reported in Fig.~\ref{fig:offset_r_rad}.}.
 
Fig.~\ref{fig:offset_r_rad} plots the $``c''$  in r band as a function
of the median cluster-centric distance of ETGs to their parent groups.
Cluster-centric   distances  are   normalised  to   the  corresponding
$R_{200}$  values,  allowing   a  meaningful  comparison  of  galaxies
residing  in  groups having  different  physical  radii.  The  $``c''$
significantly increases from about  $-8.9$ in the very central cluster
regions,  $R/R_{200}  \sim 0.1$,  to  about  $-8.865$  in the  group's
outskirts  ($R/R_{200}  \sim 7$).   The  range  of  variation is  very
similar to that found for local density, implying that both trends are
providing the same  kind of information.  It is  interesting to notice
that the  linear fits of  $``c''$ vs. $\log \Sigma_N$  and $R/R_{200}$
have  very similar  rms values,  amounting  to $0.009  \pm 0.001$  and
$0.008      \pm      0.001$,      respectively.       Notice      that
Fig.~\ref{fig:offset_r_rad} also  plots the  value of $``c''$  for the
field  sample.  To  this aim,  a fictitious  cluster-centric distance,
$R_{field}/R_{200}$, is  assigned to field galaxies,  using the linear
fit  of   $``c''$  versus  $R/R_{200}$,   in  the  same  way   as  for
$\Sigma_{field}$ (see above).

\begin{figure}
\begin{center}
\includegraphics[height=80mm]{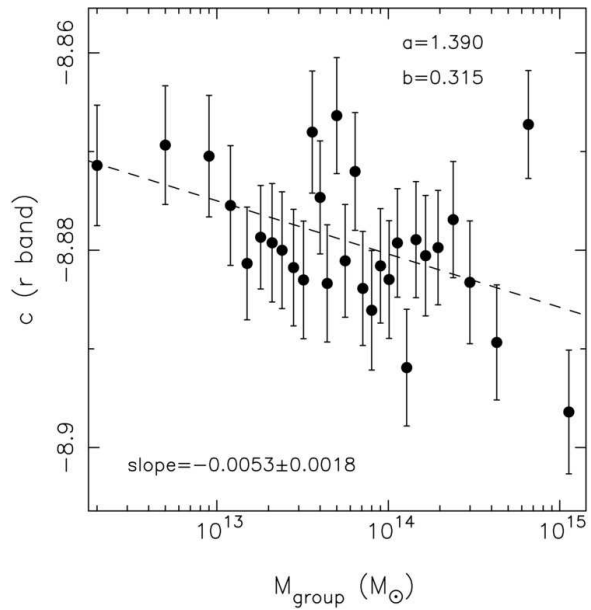}
\caption{Same as Fig.~\ref{fig:offset_r_dens}  but plotting ``c'' as a
  function of the mass of the parent group where galaxies reside.  The
  field sample is not shown in the plot.
\label{fig:offset_r_mass}
}
\end{center}
\end{figure} 

\begin{figure}
\begin{center}
\includegraphics[height=70mm]{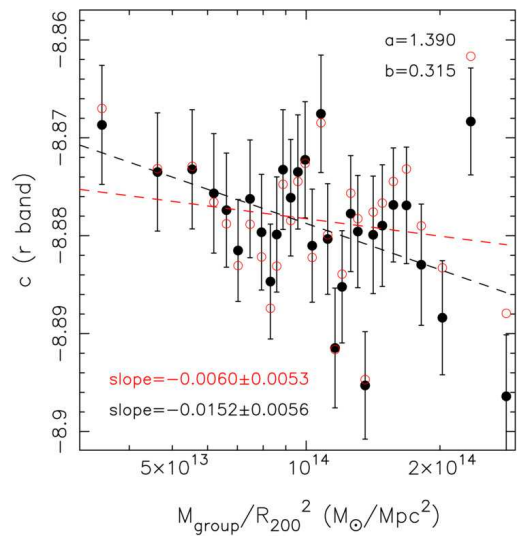}
\caption{ Same as Fig.~\ref{fig:offset_r_dens} but plotting ``c'' as a
  function  of the projected  mass density,  $M_{group}/R_{200}^2$, of
  the parent groups where ETGs  reside.  Red circles are the values of
  $``c''$ corrected  for the expected  amount of variation  of $``c''$
  among  the different  bins due  to  the different  local density  of
  galaxies in them.   The corresponding best-fit line is  shown by the
  red dashed line.  Its slope  value is reported in the lower--left of
  the plot.
\label{fig:offset_r_mrad}
}
\end{center}
\end{figure} 

Fig.~\ref{fig:offset_r_mass} plots  the $``c''$  as a function  of the
median  mass   value,  $M_{group}$,  of  the   groups  where  galaxies
reside. We find that the offset of the FP decreases as the parent halo
mass increases. The variation  is significant at the 3~$\sigma$ level,
as shown by the linear  fit of $``c''$ versus $M_{group}$, whose slope
value,  $0.053  \pm 0.018$,  is  reported  in  the same  Figure.   The
variation of $``c''$  with group mass is much  weaker than those found
for   $R/R_{200}$   and   $\Sigma_N$.    This   is   also   shown   in
Fig.~\ref{fig:offset_r_mrad}, where we plot  the $``c''$ as a function
of  the  projected  mass  density,  $M_{group}/R_{200}^2$,  of  parent
groups.  This plot can be more directly compared to the variation with
local  density $\Sigma_N$.   The FP  offset tends  to decrease  as the
$``global''$  density  increases. However,  in  the  latter case,  the
variation   is  much   weaker   than  that   with  $\Sigma_N$.   Using
Eq.~\ref{eq:c_dens},  we can  correct  the values  of  $``c''$ in  the
different  bins   of  $M_{group}/R_{200}^2$,  removing   the  expected
variation due to the different  median value of $\Sigma_N$ of galaxies
in  them.  The  corrected points  are  shown  as  red circles  in  the
Figure.  The   trend  of   $``c''$  with  $``global''$   mass  density
disappears,  proving  that  the  environmental  variation  of  the  FP
intercept is  essentially driven by $\Sigma_N$. In  other terms, local
environment  turns out  to be  the main  driver of  differences  in FP
offset.

\begin{figure*}
\begin{center}
\includegraphics[height=160mm]{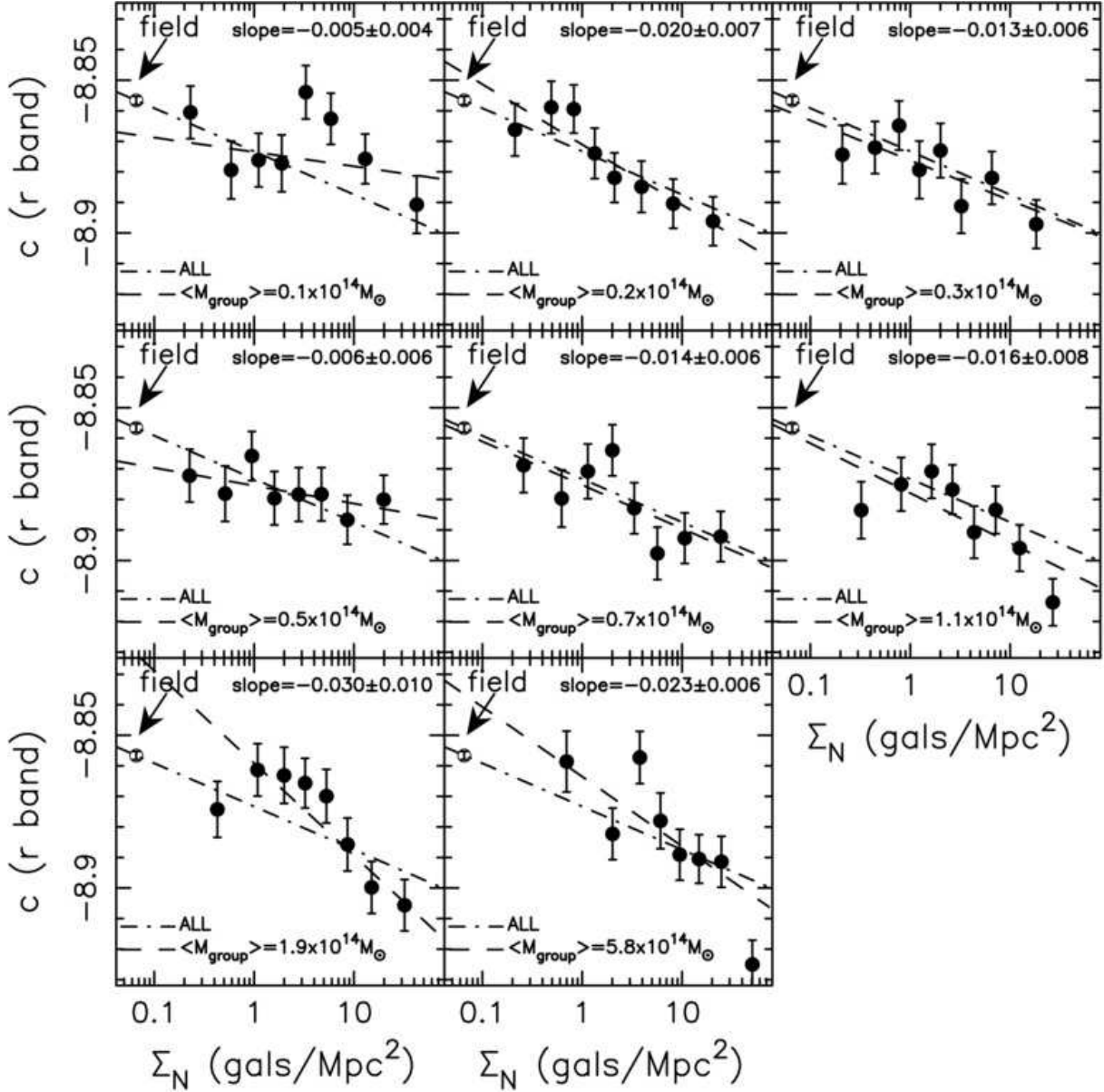}
\caption{Offset  $``c''$ of  the  FP  as a  function  of local  galaxy
  density for  different bins of parent group  mass, $M_{group}$. Each
  panel is the same  as Fig.~\ref{fig:offset_r_dens} but plotting only
  the FP  offset for ETGs in  a given bin of  $M_{group}$.  The median
  value of $M_{group}$, $< \!  M_{group} \! >$, increases from left to
  right and top to bottom, being reported in the lower--left corner of
  each  panel.  The  dashed lines  are the  linear fits  of  ``c'' vs.
  $\log \Sigma_N$, with their slope values being reported in the upper
  part  of each  panel.   The dot--dashed  line  is the  same for  all
  panels, and corresponds to the  best-fitting line of ``c'' vs. $\log
  \Sigma_N$ for the entire r-band sample. It coincides with the dashed
  black line in Fig.~\ref{fig:offset_r_dens}.
\label{fig:offset_r_mass_dens}
}
\end{center}
\end{figure*}

\begin{figure}
\begin{center}
\includegraphics[height=70mm]{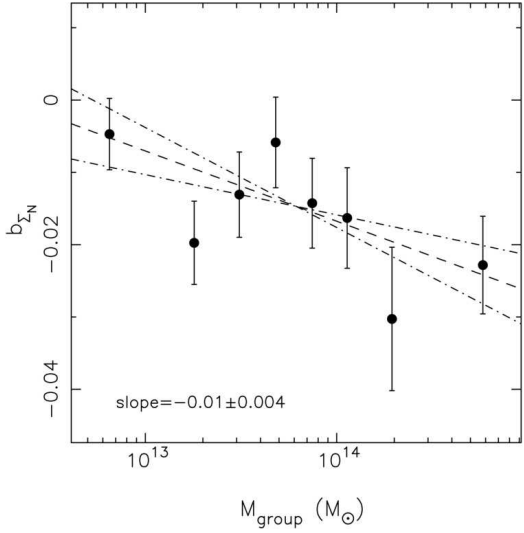}
\caption{Slope   of  the  $``c''$   vs.   $\log   \Sigma_N$  relation,
  $b_{\Sigma_N}$, as  a function of  the median mass of  parent groups
  where ETGs reside.   The slope of the best-fit  line (dashed line in
  the  plot) is reported  in the  lower--left part  of the  plot.  The
  slope  value  differs from  zero  at  the $2.5~\sigma$  significance
  level. The  dot--dashed lines  mark the $\pm  1$~$\sigma$ confidence
  levels around the best-fit.
\label{fig:bsigmaN_mass}
}
\end{center}
\end{figure}

In order  to further  analyse this point,  we derive the  variation of
$``c''$ with  density for different  bins of $M_{group}$.   The r-band
sample of ETGs is binned simultaneously with respect to $\Sigma_N$ and
$M_{group}$,  with a total  of 64  bins, each  bin including  the same
number  of 261  galaxies. Fig.~\ref{fig:offset_r_mass_dens}  plots the
$``c''$  as a  function  of  $\Sigma_N$ for  different  bins of  group
mass. The ``c'' tends to  decrease as $\Sigma_N$ increases in all mass
bins.    Each   trend   is   modelled   by   a   linear   fit   as   in
Fig.~\ref{fig:offset_r_dens}  (see Eq.~\ref{eq:c_dens}),  and compared
to the linear fit obtained for the entire sample (dot--dashed lines in
the  Figure).  The  slope value,  $b_{\Sigma_N}$, of  the  $``c''$ vs.
$\log \Sigma_N$  relation tends  to become progressively  steeper from
poor  groups to  rich  clusters.  In  fact,  the $b_{\Sigma_N}$  value
changes  from about $-0.005$  for $M_{group}  \sim 0.1  \times 10^{14}
M_{\odot}$ to  about $-0.023$ for  $M_{group} \sim 5.8  \times 10^{14}
M_{\odot}$.   For  comparison,  the  slope  of the  $``c''$  --  $\log
\Sigma_N$    relation    for    the    entire   r-band    sample    is
$b_{\Sigma_N}=-0.014             \pm            0.002$            (see
Fig.~\ref{fig:offset_r_dens}).  Fig.~\ref{fig:bsigmaN_mass}  plots the
$b_{\Sigma_N}$ as a  function of the median group  mass $M_{group}$. A
simple  linear fit  shows  that  the steepening  of  $``c''$ --  $\log
\Sigma_N$ relation with group  mass is significant at the $2.5~\sigma$
level.

\subsection{The FP offset from g through K}
\label{sec:offset_optNIR}
\begin{figure}
\begin{center}
\includegraphics[height=80mm]{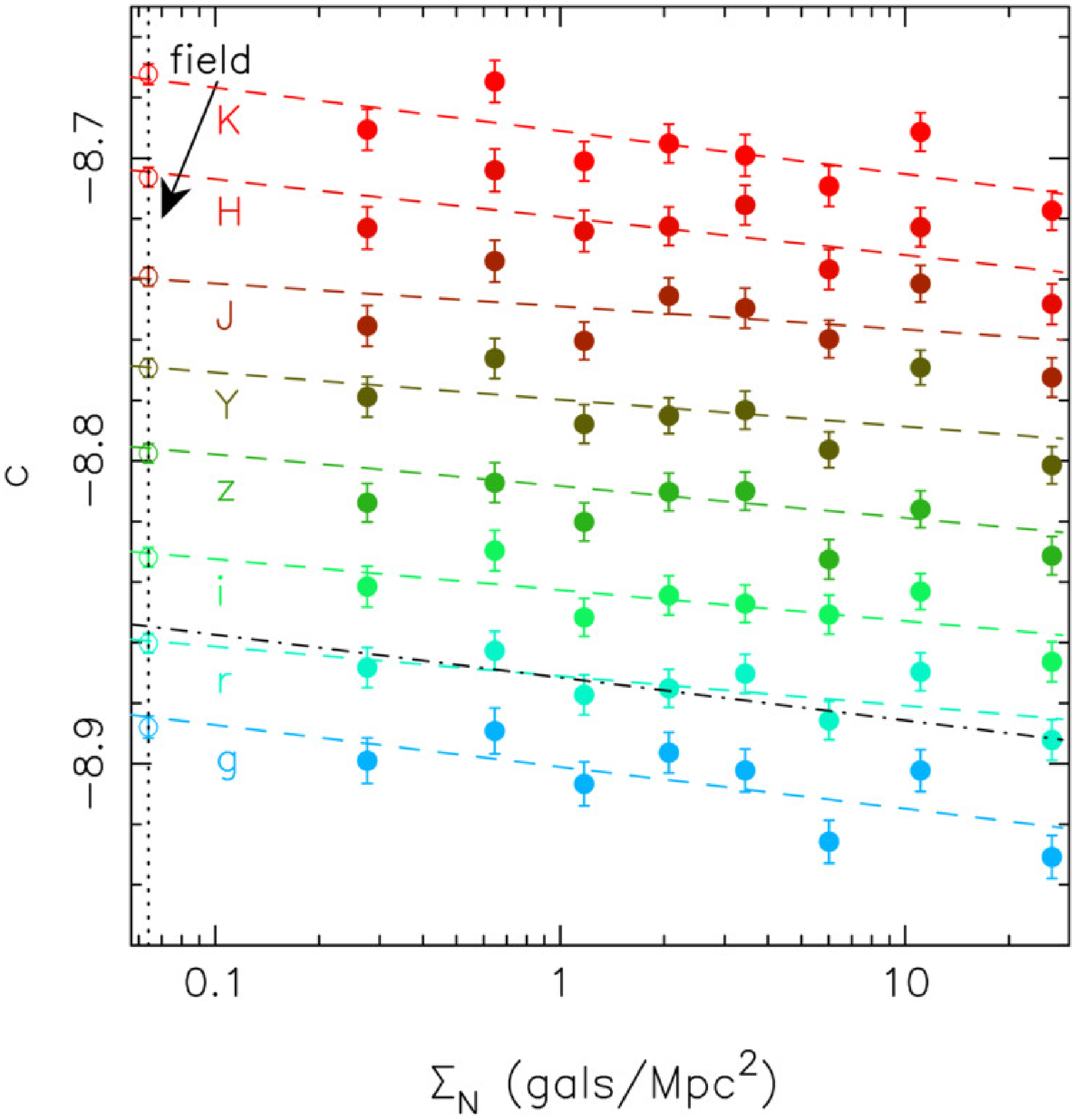}
\caption{Dependence of  the FP offset  on local galaxy density  in the
  $grizYJHK$   wavebands.   Different   wavebands  are   plotted  with
  different colours, as shown by  the labels below the dashed lines in
  the left part  of the plot.  Galaxies in  the optical+NIR samples of
  ETGs  residing in groups  are binned  according to  $\Sigma_N$, with
  each bin  including the same number  of galaxies. For  each bin, the
  $``c''$ is derived for the  same sample of galaxies from $g$ through
  $K$. The values of $``c''$  for the field sample (empty circles) are
  reported  in  the same  figure,  adopting  the  local density  value
  assigned   to    field   galaxies    from   the   linear    fit   in
  Fig.~\ref{fig:offset_r_dens}  (see Sec.~\ref{sec:offset_rband}). The
  dashed  lines are  the  linear fits  to  the trend  of $``c''$  with
  $\Sigma_   N$  in   each   band.   Error   bars  denote   $1~\sigma$
  uncertainties. Notice that suitable  shifts have been applied to the
  values  of $``c''$  in different  bands  to better  display all  the
  trends in the same plot.
\label{fig:offset_optNIR_dens}
}
\end{center}
\end{figure}

Using the optical+NIR  sample of ETGs, we derive  the variation of the
FP offset as  a function of environment, from the  $g$ through the $K$
band. We bin the optical+NIR  sample of group galaxies with respect to
$\Sigma_N$, each bin  including the same of number  of $273$ galaxies.
This results into a total of  eight bins. For a given bin, the $``c''$
is computed from Eq.~\ref{eq:c}, using  the same sample of galaxies in
all wavebands.  Fig.~\ref{fig:offset_optNIR_dens}  plots the FP offset
as  a function  of $\Sigma_N$  from $g$  through $K$.   The  values of
$``c''$ for the optical+NIR sample of field galaxies are also shown in
the Figure,  adopting the $\Sigma_N$  value ($\Sigma_{field}$) derived
for  field   galaxies  in  Sec.~\ref{sec:offset_rband}.   The  $``c''$
smoothly  decreases  from  low  to  high  density  regions,  with  all
wavebands showing the same trend. In order to characterise the trends,
we fit the data in each band with a linear relation:
\begin{equation}
c = a_{_{\Sigma_N,X}} + b_{_{\Sigma_N,X}} \log \Sigma_N,  
\label{eq:c_dens_optNIR}
\end{equation}
with $X=grizYJHK$.   The fit is  done by a least-squares  method, with
$``c''$   as  dependent  variable.    The  uncertainties   on  fitting
coefficients are  estimated by $N=500$  bootstrap iterations, shifting
the values of $``c''$  according to their uncertainties.  The best-fit
values of $a_{_{\Sigma_N,X}}$  and $b_{_{\Sigma_N,X}}$ are reported in
Tab.~\ref{tab:c_dens_grizYJHK}. Notice  that the variation  of $``c''$
per decade  in local density, i.e. the  $b_{_{\Sigma_N,X}}$, are fully
consistent,   within  the   corresponding  uncertainties,   among  the
available wavebands.  The values of $a_{_{\Sigma_N,r}}$ ($=-8.8713 \pm
0.0024$)  and $b_{_{\Sigma_N,r}}$  ($=-0.0097 \pm  0.0027$)  are fully
consistent with  those derived  for the optical  sample of  ETGs, i.e.
$a_{\Sigma_N}=  -8.8734  \pm  0.0018$  and $b_{\Sigma_N}=  -0.014  \pm
0.002$ (see  Sec.~\ref{sec:offset_rband}).  We do not  attempt here to
derive the $``c''$  as a function of both  $\Sigma_N$ and $M_{group}$.
In fact, due  to the smaller sample size of  the optical+NIR sample of
ETGs,  binning  the  $``c''$  with  respect  to  both  $\Sigma_N$  and
$M_{group}$ would overly reduce the sample size in each bin.

\begin{table}
\centering
\small
\begin{minipage}{70mm}
 \caption{Coefficients of the linear fits to the trend of $``c''$ with
   local density in different wavebands.}
  \begin{tabular}{c|c|c}
   \hline
    waveband &  $a_{_{\Sigma_N,X}}$ & $b_{_{\Sigma_N,X}}$ \\
   \hline
$g$ & $-9.1452 \pm  0.0030$ & $-0.0124 \pm  0.0034$ \\ 
$r$ & $-8.8713 \pm  0.0024$ & $-0.0097 \pm  0.0027$ \\ 
$i$ & $-8.8244 \pm  0.0025$ & $-0.0096 \pm  0.0027$ \\ 
$z$ & $-8.7134 \pm  0.0026$ & $-0.0108 \pm  0.0025$ \\ 
$Y$ & $-8.5496 \pm  0.0027$ & $-0.0098 \pm  0.0028$ \\ 
$J$ & $-8.5357 \pm  0.0026$ & $-0.0080 \pm  0.0028$ \\ 
$H$ & $-8.3709 \pm  0.0027$ & $-0.0119 \pm  0.0030$ \\ 
$K$ & $-8.2286 \pm  0.0027$ & $-0.0153 \pm  0.0030$ \\ 
   \hline
  \end{tabular}
\label{tab:c_dens_grizYJHK}
\end{minipage}
\end{table}

\section{Stellar population properties of ETGs in different environments}
\label{sec:offset_sp}
\begin{figure}
\begin{center}
\includegraphics[height=80mm]{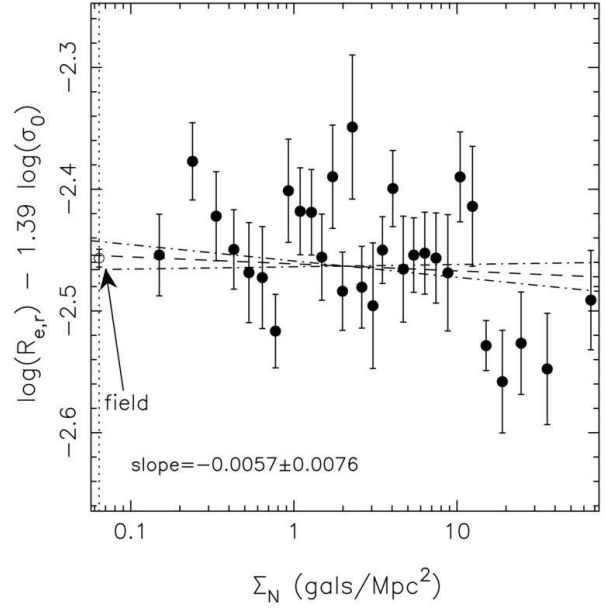}
\caption{The  peak  value  of  the   $\log  R_e  -  a  \log  \sigma_0$
  distribution, computed by the  bi-weight statistics, is plotted as a
  function of local  galaxy density for the samples  of ETGs in groups
  (filled  circles) and  the field  sample (empty  circle).  The field
  sample is marked by the  vertical dotted line. The best-fitting line
  to the data is plotted as  a dashed line. The dot--dashed lines mark
  the $\pm 1~\sigma$ confidence levels on the best-fit relation.
\label{fig:logre_r_dens}
}
\end{center}
\end{figure}

\begin{figure}
\begin{center}
\includegraphics[height=80mm]{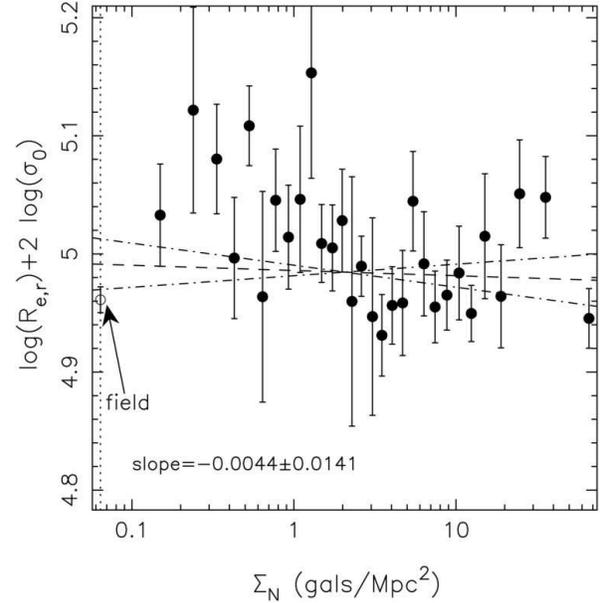}
\caption{Same  as  Fig.~\ref{fig:logre_r_dens}  but for  the  quantity
  $\log  R_{e,r} +  2  \log \sigma_0$,  which  is a  proxy for  galaxy
  dynamical mass.
\label{fig:mdyn_r_dens}
}
\end{center}
\end{figure}

Using the virial theorem, one can  rewrite the equation of the FP as a
power-law  relation between the  mass-to-light ratio  of ETGs  and two
other  variables, such as  the mass  $M$, luminosity  $L$, $\sigma_0$,
$R_e$, and  mean surface brightness  (see paper II). The  intercept of
the FP turns out to be proportional to the mean $M/L$ ratio of ETGs at
a given point of the galaxy sequence, parametrised in terms of the two
adopted variables.  Hence, the  strong variation of $``c''$ with local
environment,              (Sec.~\ref{sec:offset_rband}             and
Sec.~\ref{sec:offset_optNIR}),  can  be   actually  interpreted  as  a
variation of $M/L$ with environment.  Such a variation can be due to a
change of $M$,  $L$, or both quantities with  $\Sigma_N$.  In order to
analyse this  point, we  plot in Fig.~\ref{fig:logre_r_dens}  the peak
value  of  the  quantity $\log  R_e  -  a  \log \sigma_0$,  i.e.   the
combination of \lre \, and \ls \, that enters the FP, as a function of
local galaxy density, for the r-band sample of ETGs. The peak value is
computed by using the bi-weight estimator. We see no significant trend
with respect to $\Sigma_N$, implying  that the trend of $``c''$ is due
to  a  variation  of  mean  surface  brightness  with  local  density.
Fig.~\ref{fig:mdyn_r_dens} also plots the  quantity $\log R_e + 2 \log
\sigma_0$, which is  a proxy for the galaxy dynamical  mass, $M$, as a
function  of $\Sigma_N$.   Even in  this case,  no variation  with the
environment  is  detected.   This  suggests  that  the  $``c''$--$\log
\Sigma_N$ relation  origins from a  change in the  average luminosity,
i.e. stellar population content, of ETGs as a function of environment,
rather    than    a   variation    of    galaxy    mass   (see    also
Sec.~\ref{sec:conc}). The optical and NIR  data can help us to further
constrain the  origin of such variation in  stellar content.  Assuming
for simplicity  that the  variation of $M/L$  with environment  can be
entirely  ascribed to  age  and metallicity,  we  write the  following
equations:
\begin{eqnarray}
b_{_{\Sigma_N,X}} & = & \frac{\delta c_X}{\delta \log \Sigma_N}=  -\frac{\delta \log M/L_X}{\delta \log \Sigma_N}  \nonumber \\
        & = & -\frac{\delta \log M_{\star}/L_X}{\delta \log t} \times \frac{\delta \log t}{\delta \log \Sigma_N} + \nonumber \\
        & - & \frac{\delta \log M_{\star}/L_X}{\delta \log Z} \times \frac{\delta \log Z}{\delta \log \Sigma_N}, 
\label{eq:age_met_dens}
\end{eqnarray}
where   $M/L_X$   ($M_{\star}/L_X$)   is   the   dynamical   (stellar)
mass-to-light ratio in the  waveband $X$ ($=grizYJHK$), the quantities
$c_{t,X} =  \delta \log M_{\star}/L_X  / \delta \log t$  and $c_{Z,X}=
\delta  \log  M_{\star}/L_X  /  \delta   \log  Z  $  are  the  partial
derivatives  of  $M_{\star}/L_X$  with  respect  to the  age  $t$  and
metallicity $Z$.   Since $\delta \log M/L_X=\delta  \log M/M_{\star} +
\delta \log M_{\star}/L_X$, Eq.~\ref{eq:age_met_dens} assumes that, on
average, the quantity $\log M/M_{\star}$, i.e. the average fraction of
dark- to  stellar-matter in ETGs,  does not change  significantly with
the  environment.   {  As  discussed in  Sec.~\ref{sec:conc},  this
  assumption is supported by  the fact that the typical dynamical-mass
  of  the ETGs in our  sample  does not  change significantly  with environment
  (see Fig.~\ref{fig:mdyn_r_dens}), and a variation of $M_{\star}$, at
  fixed $M$, is unlikely.}.  The  quantities $\delta \log t / \delta
\log \Sigma_N$  and $\delta \log  Z / \delta \log  \Sigma_N$ represent
the  variation of  age  and  metallicity per  decade  in local  galaxy
density, and are the parameters we aim to constrain.  { We estimate
  the quantities $c_{t,X}$  and $c_{Z,X}$ as detailed in~\citet{LdC09}
  and  paper~II.   In  short,  we  construct  several  simple  stellar
  population (SSP) models with the ~\citet{BrC03} synthesis code.  All
  models have a Scalo IMF,  and different ages and metallicities.  Age
  values span a range of  $\sim 8$ to $\sim 13$~Gyr, while metallicity
  varies from $0.2$ to $2.5  Z_\odot$.  For each model (i.e.  each age
  and   metallicity),    we   compute   the    mass-to-light   ratios,
  $M_{\star}/L_X$,   by  folding  the   corresponding  SED   with  the
  $grizYJHK$   filter  curves.    For   each  band,   the  values   of
  $M_{\star}/L_X$ from  the different models are fitted  with an eight
  order polynomial in  $\log t$ and $\log Z$.  The  rms of the fitting
  is smaller than a few percent in all the bands.  The polynomial fits
  provide  a  simple,  analytic  tool  to  calculate  the  derivatives
  $c_{t,X}$  and  $c_{Z,X}$  for  an  SSP model  with  given  age  and
  metallicity.  In  practise, we solved  Eqs.~\ref{eq:age_met_dens} by
  evaluating  the  derivatives  at  an  age  of  $10  Gyr$  and  solar
  metallicity.  We  verified that the results are  very insensitive to
  the  choice of  such age  and metallicity  values.   Moreover, using
  different stellar population models,  as described in paper II, does
  not  change  at all  the  results.  Eqs.~\ref{eq:age_met_dens}  were
  solved  in a  least-squares  sense, using  the  estimated values  of
  $c_{t,X}$ and $c_{Z,X}$, and  the values of $b_{_{\Sigma_N,X}}$ from
  Tab.~\ref{tab:c_dens_grizYJHK}}.    In   order   to   estimate   the
uncertainties on the solutions, $\delta \log t / \delta \log \Sigma_N$
and $\delta \log Z / \delta \log \Sigma_N$, we repeat the procedure by
shifting the $b_{_{\Sigma_N,X}}$'s according to their uncertainties.
 
\begin{figure}
\begin{center}
\includegraphics[height=80mm]{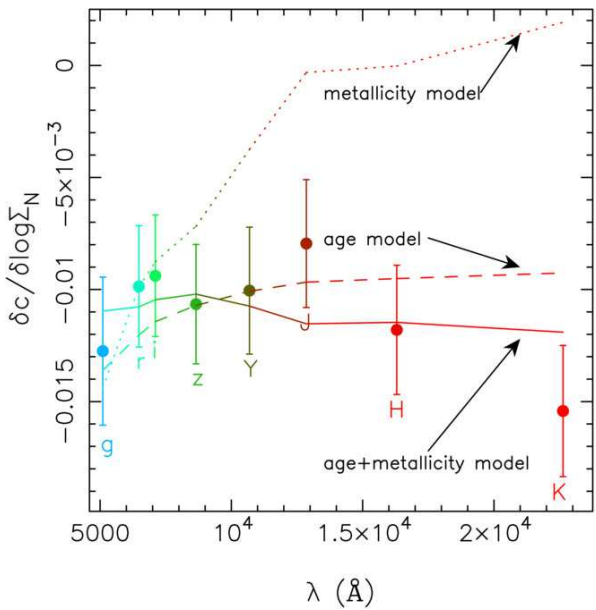}
\caption{Variation  of the  FP intercept  per decade  in  local galaxy
  density as  a function  of the effective  wavelength of  each filter
  (out  of  $grizYJHK$). The  solid,  dashed,  and  dotted curves  are
  age+metallicity, pure  age, and pure  metallicity stellar population
  models (see  the text).  Each  waveband is indicated by  a different
  colour, as shown by the labels beneath the error bars.
\label{fig:cdens_model}
}
\end{center}
\end{figure}

Fig.~\ref{fig:cdens_model} plots the $b_{_{\Sigma_N,X}}$ as a function
of the effective  wavelength of the filter $X$,  from $g$ through $K$.
The solid  curve is obtained by inserting  the best-fitting solutions,
$\delta \log  t / \delta  \log \Sigma_N$ and  $\delta \log Z  / \delta
\log  \Sigma_N$,  into  Eq.~\ref{eq:age_met_dens}, and  computing  the
expected  values  of  $b_{_{\Sigma_N,X}}$.   We see  that  a  combined
variation of age and metallicity is  able to match very well the value
of $b_{_{\Sigma_N,X}}$ in all wavebands.  The dashed and dotted curves
in the Figure  are obtained with the same  procedure by setting either
$\delta \log  Z / \delta \log  \Sigma_N=0$ or $\delta \log  t / \delta
\log \Sigma_N=0$ in Eq.~\ref{eq:age_met_dens}. They correspond to pure
age  and metallicity  models of  the variation  of $``c''$  with local
density. The  pure metallicity model is clearly  inconsistent with the
observations.   In fact,  a difference  in metallicity  would  tend to
produce a negligible difference  of mass-to-light ratios in K-band, as
the  NIR light  is  very insensitive  to  differences in  metallicity,
through line-blanketing. On the other  hand, a pure age model is still
able to match the observations, but with a large discrepancy, of $\sim
2~\sigma$, in K-band.

\begin{figure}
\begin{center}
\includegraphics[height=80mm]{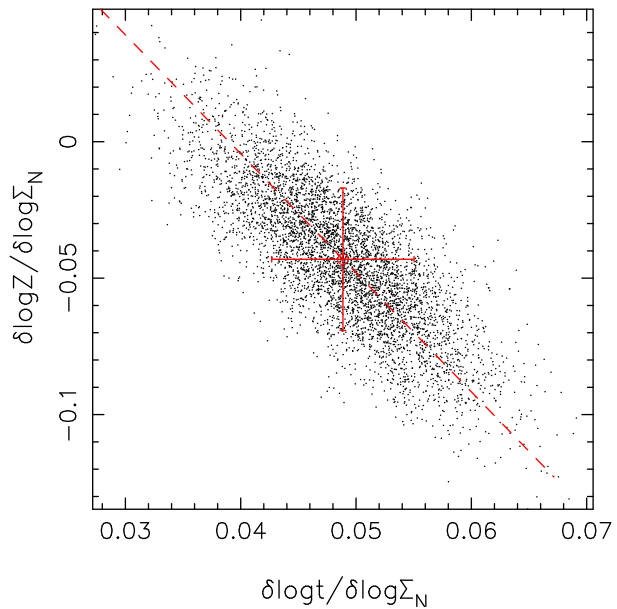}
\caption{Variations of age and  metallicity per decade in local galaxy
  density, $\delta \log t / \delta \log \Sigma_N$ and $\delta \log Z /
  \delta \log  \Sigma_N$, implied  by the variation  of the  FP offset
  with $\log \Sigma_N$ in the different wavebands.  The best solutions
  of Eq.~\ref{eq:age_met_dens} are marked by the red circle. The black
  points are the values of $\delta  \log t / \delta \log \Sigma_N$ and
  $\delta  \log Z  / \delta  \log \Sigma_N$  obtained by  shifting the
  $b_{_{\Sigma_N,X}}$  according to  the  corresponding uncertainties.
  {  The  dashed line  is  obtained  by  an ordinary  least-squares
    fitting procedure,  taking the  arithmetic average of  the results
    obtained by  changing the  role of the  dependent variable  in the
    fitting. }
\label{fig:dage_dmet_c}
}
\end{center}
\end{figure}

Fig.~\ref{fig:dage_dmet_c} plots the values of $\delta \log t / \delta
\log \Sigma_N$  vs.  $\delta \log  Z / \delta \log  \Sigma_N$ obtained
from  the procedure  described above.   { The  scatter in  the plot
  (black   dots)  reflects  the   measurement  uncertainties   on  the
  quantities  $b_{_{\Sigma_N,X}}$.   As expected  from  the fact  that
  $\delta \log Z / \delta \log  \Sigma_N $ and $\delta \log t / \delta
  \log  \Sigma_N$ are  estimated  from the  same  system of  equations
  (Eqs.~\ref{eq:age_met_dens}),    the     measurement    errors    on
  $b_{_{\Sigma_N,X}}$ imply a correlated variation of $\delta \log Z /
  \delta \log \Sigma_N  $ and $\delta \log t  / \delta \log \Sigma_N$.
  A linear fitting of the $\delta  \log Z / \delta \log \Sigma_N $ and
  $\delta \log t / \delta  \log \Sigma_N$ values (dashed line) gives a
  slope of  $-4.36 \pm  0.05$.   Taking the  mean and  standard
  deviation  of  $\delta \log  Z /  \delta \log  \Sigma_N  $'s and
  $\delta \log t / \delta \log \Sigma_N$'s, we obtain} `$\delta \log Z
/ \delta \log \Sigma_N =-0.043 \pm  0.023$ and $\delta \log t / \delta
\log \Sigma_N = 0.048 \pm 0.006$, i.e.  ETGs in higher density regions
are  older, by  $\sim  11\%$ per  decade  in local  density, and  less
metal-rich than galaxies at low  density.  We notice that the value of
$\delta \log  Z / \delta  \log \Sigma_N $  is consistent with  zero at
about $2~\sigma$,  implying that, as  noticed above, a pure  age model
might be able  by itself to explain the observations.   For a pure age
model, the  average value  of $\delta \log  t / \delta  \log \Sigma_N$
reduces to $0.04 \pm 0.004$, rather than $0.048 \pm 0.006$.

In Sec.~\ref{sec:offset_fp},  we have reported some  evidence that the
variation  of $``c''$  per decade  in  local density  is stronger  for
galaxies  residing in  more massive  (relative to  poor)  clusters. As
shown in  Fig.~\ref{fig:bsigmaN_mass}, for $M_{group}  \widetilde{>} 2
\times 10^{14}  M_\odot$, the  absolute value of  $b_{_{\Sigma_N}}$ is
about   twice   larger   than    that   of   $-0.0097$   reported   in
Tab.~\ref{tab:c_dens_grizYJHK}.   For   a   pure   age   model,   from
Eq.~\ref{eq:age_met_dens},  one  would infer  that,  for more  massive
clusters, the variation  in age per decade of  local galaxy density is
around twice  larger than that reported  above, i.e. $\delta  \log t /
\delta \log \Sigma_N \sim 0.08$.

\section{Slopes of the FP as a function of the environment}
\label{sec:slopes_fp}
So far  we have analysed  the offset of  the FP, interpreting it  as a
difference  of the  average mass-to-light  ratio of  ETGs  residing in
different environments.   In this section, we study  the dependence of
$``a''$    and     $``b''$    themselves    on     environment.     In
Sec.~\ref{sec:slopes_rband}, we analyse the slopes $``a''$ and $``b''$
in the r band, taking advantage of the large size of the optical ETG's
sample.   In  Sec.~\ref{sec:slopes_optNIR},  we  derive  the  waveband
dependence of FP slopes.

\subsection{The slopes of the FP in r band}
\label{sec:slopes_rband}
We start  comparing the  slopes of  the FP for  the r-band  samples of
field and group galaxies (Sec.~\ref{sec:field_group_samples}). To this
aim, we have to account  for two different effects, (i) field galaxies
have on average  lower mass-to-light ratios than those  in groups (see
Sec.~\ref{sec:offset_sp}),  and  (ii)  ETGs  are  characterized  by  a
luminosity segregation,  for which  brighter ({ i.e.   higher $L$})
galaxies mostly  inhabit the  denser cluster regions.   { Moreover,
  not  only $L$, but  also galaxy  radii are  expected to  change with
  environment,  as   they  correlate  with   position  within  cluster
  (e.g.~Cypriano et al.   2006)}.  Point (i) implies that,  at a given
mass, an ETG  in a lower density region is on  average brighter than a
galaxy with  the same  mass, but in  a denser  environment.  Moreover,
even  removing  this  effect,  the  relative  distribution  of  galaxy
luminosities and mean surface brightnesses might depend on environment
(point  ii).   To account  for  points (i)  and  (ii),  we proceed  as
follows.    First,   we   fix   the   slopes  of   the   FP,   as   in
Sec.~\ref{sec:offset_rband}, and  derive the intercepts of  the FP for
the  field  and  group  samples.   From  Eq.~\ref{eq:c},  the  average
luminosity difference between the two  samples is given by $\delta E =
-(\delta c)/b  = -0.077  \pm 0.007$~mag.  Then,  we shift the  \mie \,
values of  galaxies in the field  sample by $-\delta E$,  and bin both
the field and  group samples with respect to  total magnitude and mean
surface brightness. For each bin,  we randomly extract the same number
of field and  group galaxies.  This procedure, described  in detail in
paper  II (see the  app.~B), allows  us to  extract two  subsamples of
field and group galaxies having  the same distribution in the space of
the effective parameters.  By construction, both subsamples consist of
the  same number  of galaxies,  i.e.  $N=11,024$  out of  $11,824$ and
$16,717$ ETGs in the field  and groups, respectively.  It follows that
any difference in the FP slopes for these two subsamples arises from
the different ways velocity dispersions are related to the effective
parameters, rather than from any spurious ($``geometric''$) effect due
to different relative fraction of galaxies in different regions of the
parameter  space.  {  In  other  words,  we  are  ``removing''  the
  dependence of galaxy luminosities and radii on environment, in order
  to measuring the genuine dependence of the FP on environment.}

\begin{figure}
\begin{center}
\includegraphics[height=80mm]{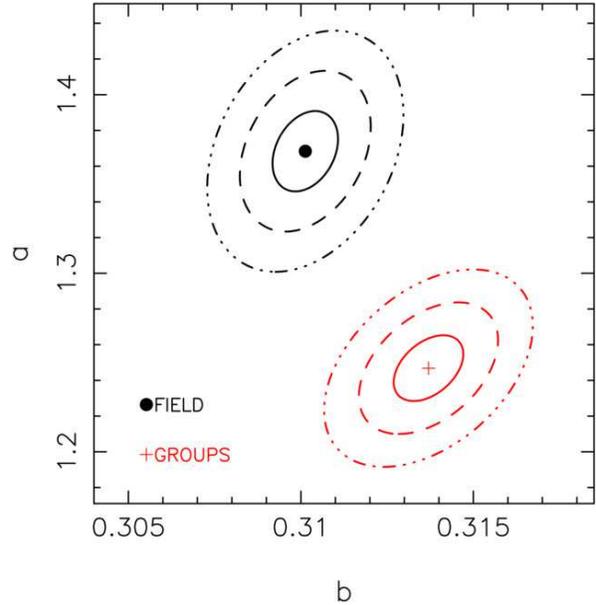}
\caption{Slopes $``a''$ and $``b''$ of the FP for field (black circle)
  and group  (red cross) galaxies. The solid,  dashed, and dot--dashed
  ellipses  mark  the $1,~2$  and  $3~\sigma$  confidence contours  on
  $``a''$ and  $``b''$, respectively. Notice  that both the  field and
  group  samples  have the  same  distribution  in  the space  of  the
  effective parameters (after  the average difference in mass-to-light
  ratio is  accounted for),  and include the  same number  of $11,024$
  ETGs (see the text).
\label{fig:FP_slopes_field_group}
}
\end{center}
\end{figure}

Fig.~\ref{fig:FP_slopes_field_group}  plots  the  $``a''$ and  $``b''$
values  for the  field and  group  subsamples. We  find a  significant
difference  between the  FP of  field and  group galaxies,  with field
galaxies  having  higher $``a''$  and  lower  $``b''$  value. For  the
$''a''$,  the difference  is significant  at $\sim  3~\sigma$.  Notice
that had we  not homogenised the luminosity and  \mie \, distributions
of  galaxies  in  the two  samples,  we  would  have found  an  almost
negligible difference in $``a''$ (at  $\sim 1 \sigma$), and a stronger
difference in $``b''$  (at $\sim 5~\sigma$). { The
  dependence of galaxy luminosities  and radii on environment plays an
  important role. If not accounted for, it would lead to underestimating
  the dependence of the coefficient $''a''$ on environment. To further
  test this  result, we have implemented  a procedure to  wash out the
  dependence of FP on environment and measure only the effect that the
  environmental dependence of galaxy luminosities and radii have on FP
  coefficients.  For  a given galaxy parameter  (e.g.  luminosity), we
  bin  the  sample  of  field+group  galaxies  with  respect  to  that
  parameter, each  bin including the same number  of $N=100$ galaxies.
  Then, we  scramble the galaxy  environments, preserving
  the number  of objects residing  in each environmental  bin (i.e.
  field and groups). The resulting shuffled sample is split again into
  field and  group galaxies. Since both galaxy  luminosities and radii
  are known to  be affected from environments (see  above), we perform
  the scrambling with respect to both $L$ and $r_e$. Also, we consider
  the case of  scrambling \mie \, (as it reflects  both $L$ and $r_e$)
  and the remaining  FP variable, \ls. The FP  slopes of the scrambled
  samples   of   field   and    group   galaxies   are   reported   in
  Tab.~\ref{tab:scrable_cof}.    The  coefficient  $''b''$   is  fully
  consistent  between  the  scrambled   samples  of  field  and  group
  galaxies.  For the $''a''$, we  see that scrambling \mie \, and \ls,
  does not impact at all  its dependence on environment. On the other
  hand, scrambling $L$ and $r_e$ has the net effect of decreasing ($by
  \sim 5 \%$) the $''a''$ for field (relative to group) galaxies. This
  result  is  fully  consistent  with  our  conclusion  above.   Field
  galaxies  have  higher $''a''$  than  group  galaxies. However,  the
  dependence  of $L$  and  $r_e$  on environment  does  not make  this
  difference  immediately  detectable, as  it  tends  to decrease
  $''a''$ for the field sample.}

\begin{table*}
\centering
\small
\begin{minipage}{104mm}
 \caption{  FP slopes  of  ETGs  in groups  and  field obtained  by
   scrambling  galaxy environment at  fixed size,  surface brightness,
   velocity dispersion, and logarithmic luminosity.}
  \begin{tabular}{c|c|c|c|c}
   \hline
 scrambling & \multicolumn{2}{c}{field} & \multicolumn{2}{c}{groups}  \\
 & $a$ & $b$ & $a$ & $b$  \\
   \hline
$r_e$ & $  1.350 \pm   0.019$ & $  0.315 \pm   0.001$ &  $  1.406 \pm   0.014$  & $   0.316 \pm   0.001$ \\ 
\mie & $  1.399 \pm   0.016$ & $  0.315 \pm   0.001$ &  $  1.388 \pm   0.016$  & $   0.316 \pm   0.001$ \\ 
\ls  & $  1.401 \pm   0.017$ & $  0.315 \pm   0.001$ &  $  1.391 \pm   0.014$  & $   0.316 \pm   0.001$ \\ 
$\log L$  & $  1.353 \pm   0.020$ & $  0.314 \pm   0.001$ &  $  1.402 \pm   0.014$  & $   0.316 \pm   0.001$ \\ 
 \hline
  \end{tabular}
\label{tab:scrable_cof}
\end{minipage}
\end{table*}

\begin{figure}
\begin{center}
\includegraphics[height=80mm]{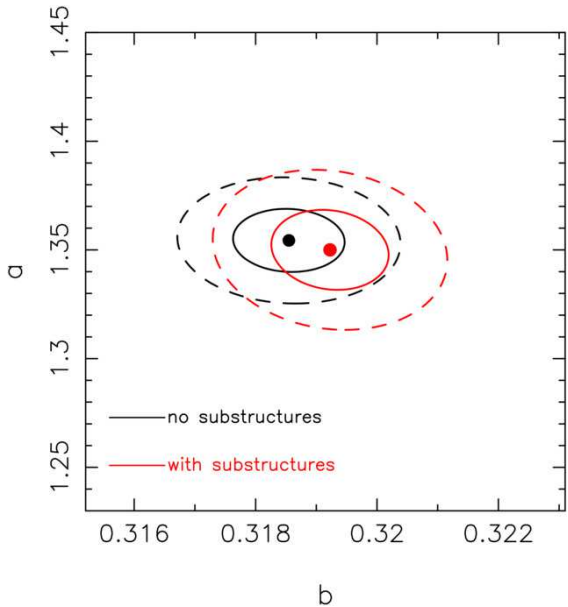}
\caption{Slopes $``a''$ and $``b''$ of the FP for galaxies residing in
  groups with (red) and  without (black) substructures.  The solid and
  dashed ellipses  mark the $1$ and $2~\sigma$  confidence contours on
  $``a''$ and $``b''$, respectively.
\label{fig:FP_slopes_r_substr}
}
\end{center}
\end{figure}

We derive  the FP slopes of  ETGs residing in groups  with and without
substructures. The presence of  substructures is characterized by a 3D
$\Delta$  test (see  Sec.~\ref{sec:substructures}). About  $44  \%$ of
groups show evidence of substructures.  ETGs in the optical sample are
first split in two subsamples  according to the presence ($N=9133$) or
not  ($N=7584$) of substructures  in the  parent groups.   Notice that
although less than  half of the groups have  substructures, there is a
large  number of  ETGs residing  in groups  with  substructures. Then,
proceeding as  described above, we extract two  subsamples of galaxies
with  the same  distribution  in the  space  of effective  parameters.
These subsamples include $7,716$  galaxies each. No correction for the
average mass-to-light ratio is applied, as we obtain similar values of
the  FP offset for  the two  samples, i.e.   $c=-8.876 \pm  0.002$ and
$c=-8.882  \pm 0.002$  for the  case with  and  without substructures,
respectively.    Fig.~\ref{fig:FP_slopes_r_substr}  compares   the  FP
slopes  for the two  $``homogenised``$ subsamples.   The corresponding
values of $''a``$ and $''b``$ are fully consistent.

\begin{figure}
\begin{center}
\includegraphics[height=95mm]{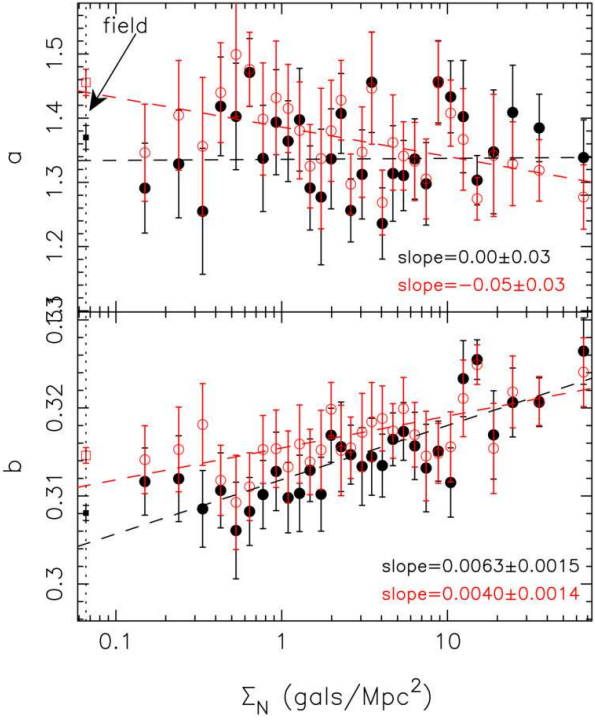}
\caption{The slopes  $``a''$ and  $``b''$ of the  FP are plotted  as a
  function  of local  galaxy density,  $\Sigma_N$.  Error  bars denote
  $1~\sigma$ uncertainties. Red colour shows the case where the slopes
  are  corrected  for  the  environmental dependence  of  the  average
  mass-to-light ratio  of ETGs and luminosity  segregation of galaxies
  in different environments  (see the text). The dashed  lines are the
  linear least-squares  fits to the  data points.  The squares  in the
  left part of the plots are  the slope's values for the field sample,
  as also indicated by the vertical dotted line and the arrow.
\label{fig:FP_DENS}
}
\end{center}
\end{figure}

To compare  the FP  slopes of galaxies  residing in different  bins of
local  density,   we  apply  the   same  binning  procedure   used  to
characterise     the    trend     of    $``c''$     with    $\Sigma_N$
(Fig.~\ref{fig:offset_r_dens}).   To derive  the slopes  of the  FP in
each bin,  we proceed in a similar  way as in the  comparison of field
and group  galaxies, i.e.   we account for  (i) the  different average
mass-to-light ratio of ETGs  in different environmental bins, and (ii)
differences in the  relative distribution of galaxies in  the space of
effective  parameters.   First,  we  estimate the  average  luminosity
difference of galaxies residing in  different bins with respect to the
field sample, using  the linear fit of $``c''$  versus $\log \Sigma_N$
(Eq.~\ref{eq:c_dens}).   For a  given  density bin,  $i$, the  average
luminosity difference is given by
\begin{equation}
 \delta E_i = - \frac{ b_{_{\Sigma_N}} \times (\log \Sigma_N - \log \Sigma_{field} )  }{b},
\end{equation}
where $\Sigma_{field}$  is the local  density value assigned  to field
galaxies  (Sec.~\ref{sec:offset_rband}).  We  then shift  the  \mie \,
values  of the  entire optical  sample of  ETGs by  $\delta  E_i$, and
extract a subsample  of galaxies with the same  luminosity and \mie \,
distributions as galaxies in the $i$-th $\Sigma_N$ bin. We compute the
relative variations, $\delta a /a$  and $\delta b/b$, of the FP slopes
of the entire  sample after these selections are  applied. The factors
$\delta  a /a$  and  $\delta b/b$  are  applied to  the  FP slopes  of
galaxies  in  the  given  local  density bin.   By  construction,  the
correction vanishes  for the  field sample, while  it become  more and
more important as  $\Sigma_N$ increases.  Fig.~\ref{fig:FP_DENS} plots
the r-band slopes of the FP  as a function of local galaxy density. In
order  to   characterise  the  trend  of  $``a''$   and  $``b''$  with
$\Sigma_N$, linear fits of both slopes with respect to $\log \Sigma_N$
are performed.  Notice that  the field sample  is not included  in the
fit.

\begin{figure}
\begin{center}
\includegraphics[height=95mm]{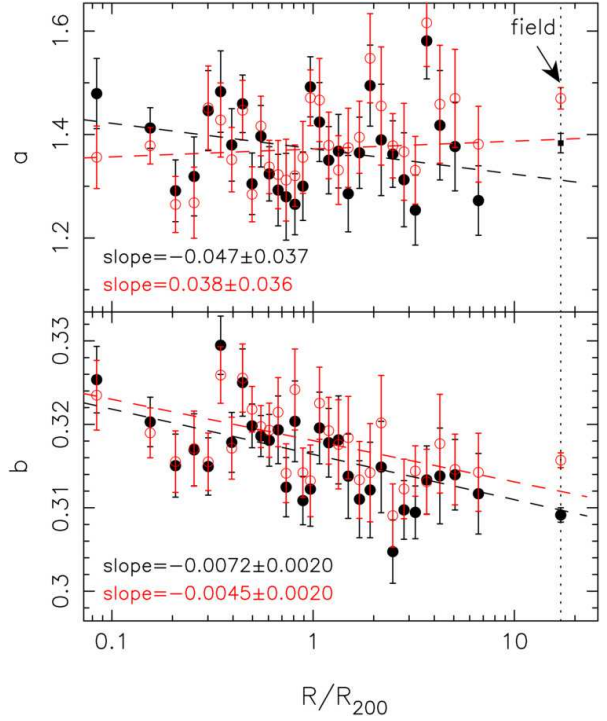}
\caption{The same  as Fig.~\ref{fig:FP_DENS} but  plotting the $``a''$
  and $``b''$ as  a function of the normalised  distance to the centre
  of the parent groups where galaxies reside.
\label{fig:FP_RADIUS}
}
\end{center}
\end{figure}

In general,  one can  notice that  the FP slopes  exhibit only  a weak
dependence on environment. Before applying any correction, the $``a''$
does not  show any  variation with $\Sigma_N$,  with the slope  of the
linear regression ($0.00  \pm 0.03$; see Fig.~\ref{fig:FP_DENS}) being
fully consistent with zero.  After the corrections are applied, we are
able to detect a $2.5~\sigma$ tendency of $``a''$ to decrease from low
to high density environments.  One  can notice that, in agreement with
this trend,  the corrected  value of $``a''$  for the field  sample is
larger  than all  the slope  values  obtained for  the higher  density
subsamples ($\Sigma_N > 1  gals/Mpc^2$).  This is consistent with what
found above, when comparing the  FP slopes of field and group galaxies
(Fig.~\ref{fig:FP_slopes_field_group}).   Notice  that  the values  of
$``a''$ and $``b''$ in Fig.~\ref{fig:FP_slopes_field_group} refer to a
different sample of field galaxies  with respect to that considered in
Fig.~\ref{fig:FP_DENS}.             In            fact,            for
Fig.~\ref{fig:FP_slopes_field_group},  two  subsamples  of  field  and
group galaxies are extracted, to match the corresponding distributions
of galaxies in the space  of structural parameters. For the comparison
in  Fig.~\ref{fig:FP_DENS}, the  FP slopes  are derived  by  using all
galaxies  in  the field  sample.   This  explains  the difference  (in
absolute  value)  of the  values  of  $``a''$  and $``b''$  for  field
galaxies      between     Fig.~\ref{fig:FP_slopes_field_group}     and
Fig.~\ref{fig:FP_DENS}.  Despite  that, the relative  difference of FP
coefficients between the field and  group samples are detected in both
cases.  The  coefficient $``b''$ tends  to increase with  local galaxy
density.   The  correction  procedure   shows  that  the  tendency  is
significant  at $\sim  3 \sigma$,  and is  mostly due  to  galaxies in
higher  density  regions  ($\Sigma_N   \widetilde{>}  10  \,  gals  \,
Mpc^{-2}$).   As  shown  in  Fig.~\ref{fig:FP_RADIUS}, the  trends  of
$``a''$ and $``b''$ with environment become weaker when plotting their
value  as  a  function  of  the  normalised  cluster-centric  distance
$R/R_{200}$.   In this case,  the corrections  are estimated  with the
same  approach described  above,  using the  fictitious  value of  the
cluster-centric distance of the field sample ($R_{field}/R_{200}$) and
the    linear    fit    of    $``c''$    versus    $R/R_{200}$    (see
Fig.~\ref{fig:offset_r_rad}).  We find that the $``b''$ increases from
the centre  to the  periphery of galaxy  groups, as expected  from the
variation of $``b''$ with local galaxy density. For the $``a''$, we do
not detect any significant variation with $R/R_{200}$.

\begin{figure}
\begin{center}
\includegraphics[height=95mm]{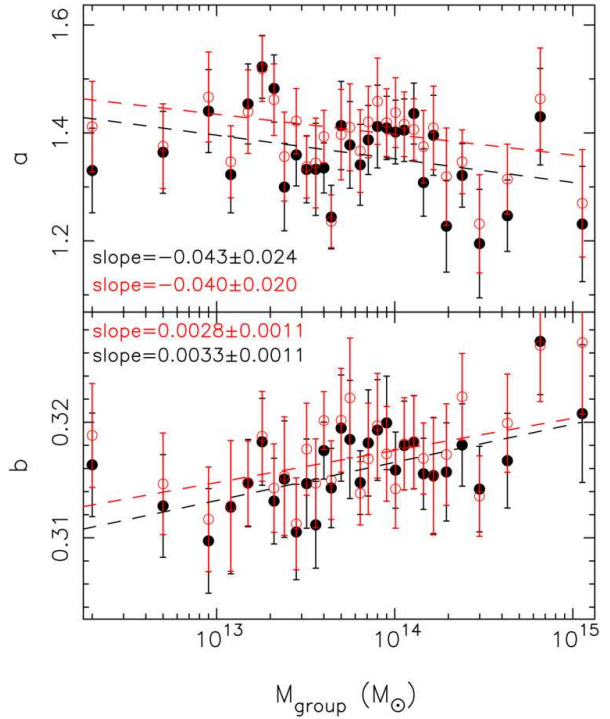}
\caption{The same  as Fig.~\ref{fig:FP_DENS} but  plotting the $``a''$
  and $``b''$  as a function of  mass of parent  groups where galaxies
  reside.
\label{fig:FP_MASS}
}
\end{center}
\end{figure} 

\begin{figure}
\begin{center}
\includegraphics[height=95mm]{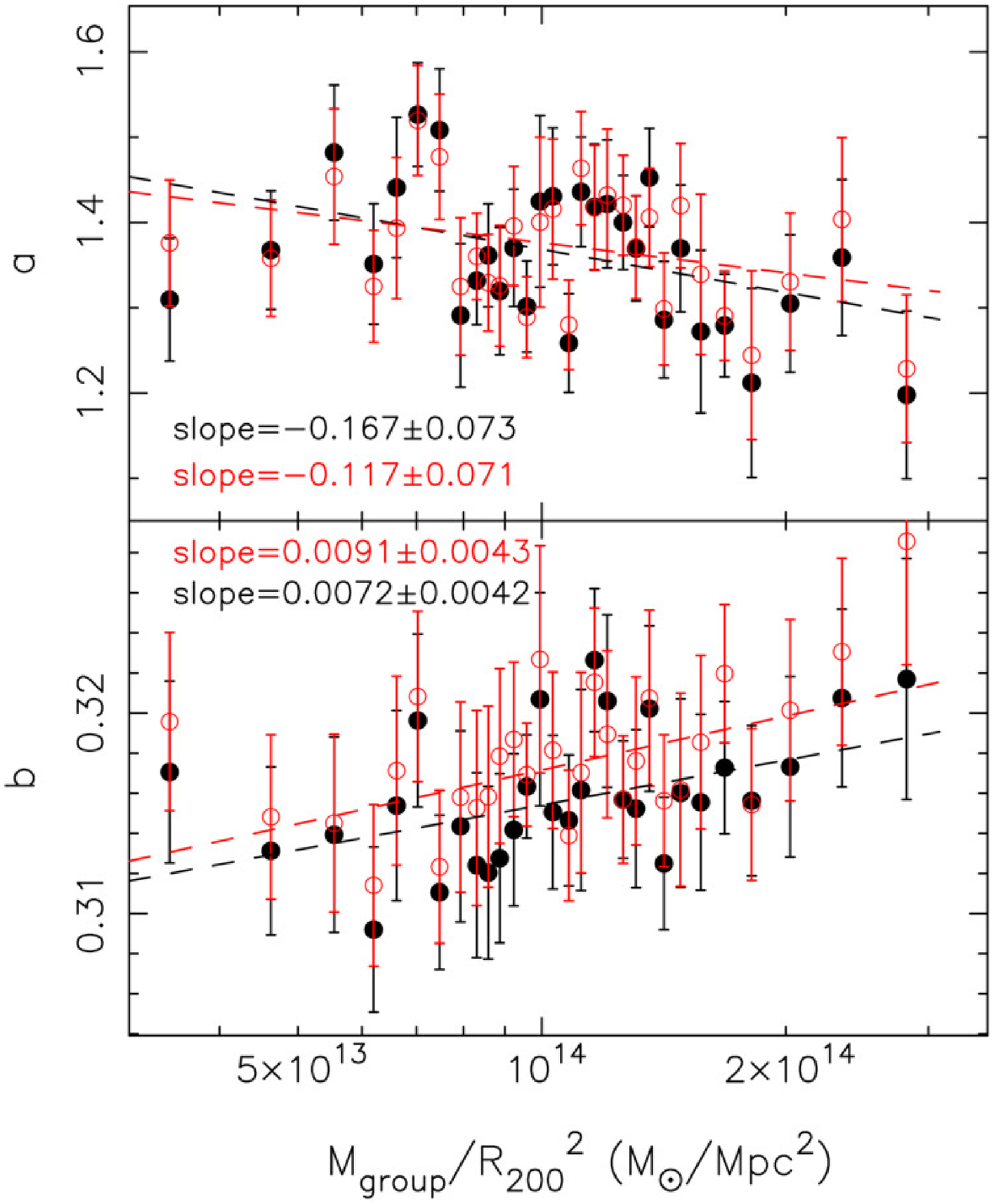}
\caption{The same  as Fig.~\ref{fig:FP_DENS} but  plotting the $``a''$
  and  $``b''$ as  a function  of $``global''$  surface  mass density,
  $M_{group}/R_{200}^2$, of the parent groups where galaxies reside.
\label{fig:FP_MRAD}
}
\end{center}
\end{figure} 

Fig.~\ref{fig:FP_MASS} plots the FP slopes  in r band as a function of
the  mass  of  the  parent  galaxy  groups.   As  for  $\Sigma_N$  and
$R/R_{200}$,  we plot  the  values  of $``a''$  and  $``b''$ with  and
without the corrections for the  effects (i) and (ii) mentioned above.
We  find that  the coefficient  $``a''$  tends to  decrease (at  $\sim
2~\sigma$) for  galaxies residing in more massive  clusters, while the
$``b''$ shows  an opposite and  more significant (at  $\sim 3~\sigma$)
trend, increasing for higher $M_{group}$. A similar result is obtained
when   characterising  the   global  environment   in  terms   of  the
$``global''$  projected mass  density,  $M_{group}/R_{200}^2$, of  the
parent galaxy groups. The $``a''$ ($``b''$) exhibits only a weak trend
to      decrease       (increase)      with      $M_{group}/R_{200}^2$
(Fig.~\ref{fig:FP_MRAD}).  As shown by  the slope values of the linear
fits, the trends of $``a''$  and $``b''$ with $``global''$ density are
only  marginally  significant,  at   less  than  $2~\sigma$.   On  the
contrary,  the   dependence  on  local   density  seems  to   be  more
significant,  as  shown  by  the  uncertainties on  the  slope  values
reported in Fig.~\ref{fig:FP_DENS}. This implies that the local rather
than the  global environment is  the main driver of  the environmental
dependence of the slopes of the FP.  The same result has been shown to
hold for the offset of the FP (Sec.~\ref{sec:offset_rband}).

\begin{figure*}
\begin{center}
\includegraphics[height=140mm]{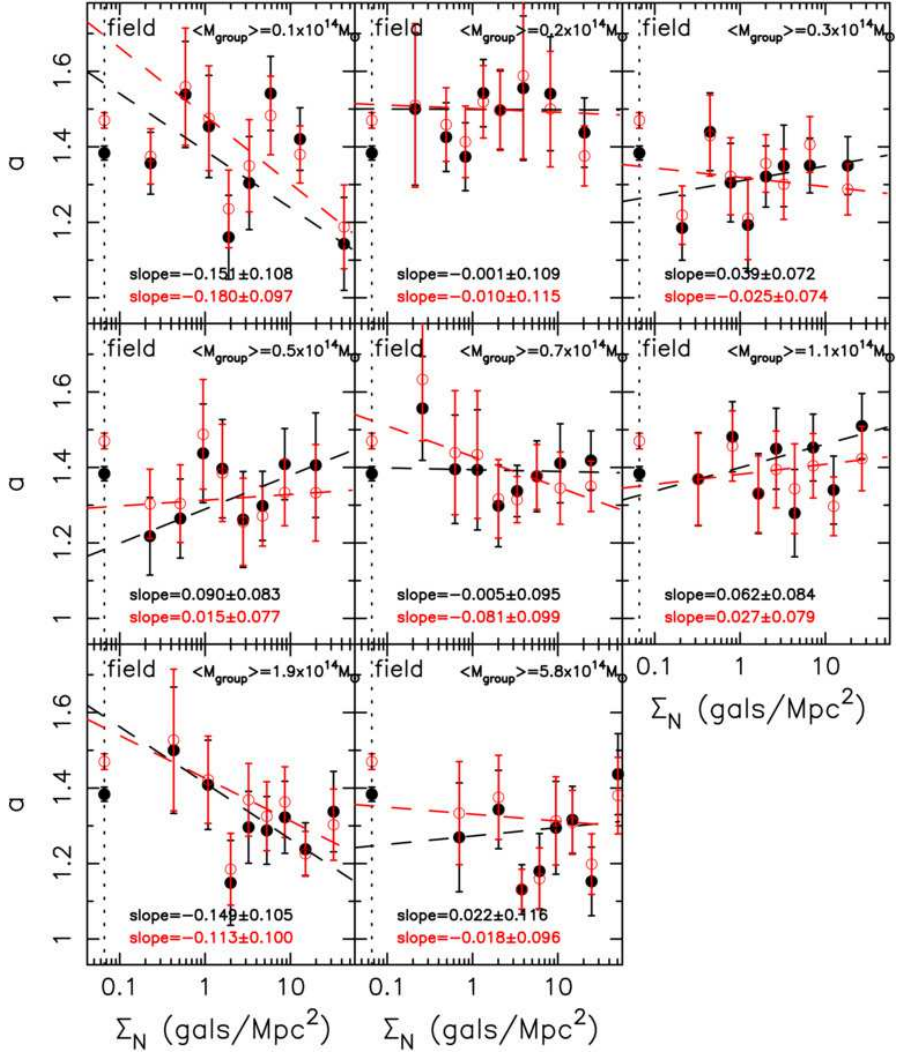}
\caption{Dependence of  the slope $``a''$ of  the FP as  a function of
  local density, $\Sigma_N$,  in different bins of mass  of the parent
  galaxy group.  Red colour  refers to the  values of the  slope after
  correcting for  differences in the FP intercept  and distribution in
  the space of structural parameters (see the text). Error bars denote
  $1~\sigma$ uncertainties.  For  each panel, the corresponding median
  value  of the  mass of  the parent  galaxy group,  $<M_{group}>$, is
  reported in  the upper-right. The $<M_{group}>$  increases from left
  to right,  and top to  bottom. The values  of $``a''$ for  the field
  sample are  reported in each panel,  and are marked  by the vertical
  dashed line. The  dashed lines are best-fit linear  relations to the
  data points,  with the corresponding slope values  being reported in
  the lower-left part of each  panel.  Notice that the $\Sigma_N$ bins
  are the same as in Fig.~\ref{fig:offset_r_mass_dens}.
\label{fig:FP_MASS_DENS_a}
}
\end{center}
\end{figure*}

\begin{figure*}
\begin{center}
\includegraphics[height=140mm]{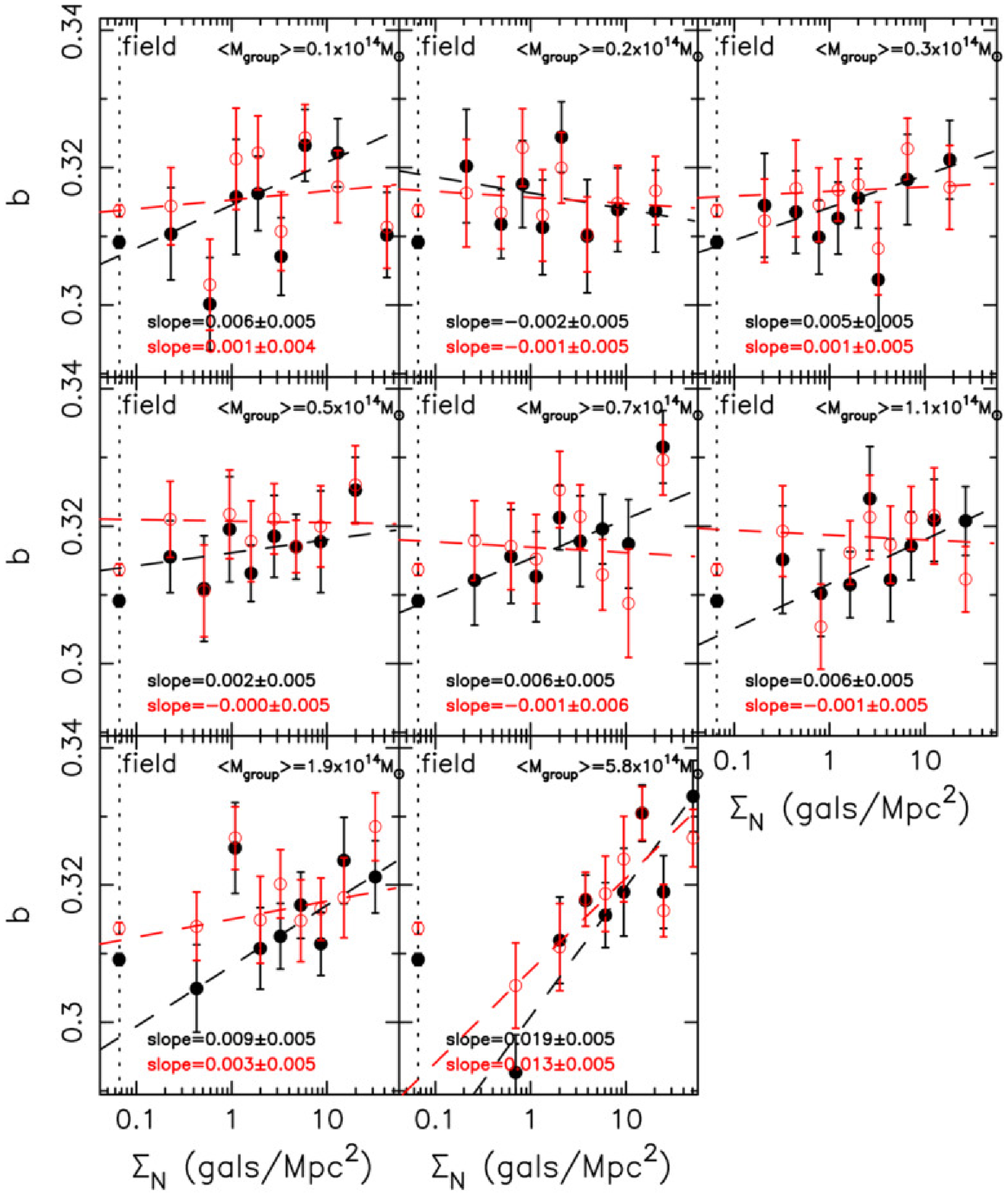}
\caption{The  same   as  Fig.~\ref{fig:FP_MASS_DENS_a}  but   for  the
  coefficient $``b''$ of the FP.
\label{fig:FP_MASS_DENS_b}
}
\end{center}
\end{figure*}

As for the $``c''$, we also analyse how the slopes of the FP depend on
local galaxy  density in different bins  of mass of  the parent galaxy
groups.   Figs.~\ref{fig:FP_MASS_DENS_a}  and~\ref{fig:FP_MASS_DENS_b}
plot  the $``a''$ and  $``b''$ as  a function  of $\log  \Sigma_N$, in
different bins  of $M_{group}$. The  binning is performed in  the same
way  as for  Fig.~\ref{fig:offset_r_mass_dens}. The  slope  values are
corrected as  in Fig.~\ref{fig:FP_DENS}. Fig.~\ref{fig:FP_MASS_DENS_a}
does not  show any significant  dependence of $``a''$ on  local galaxy
density for all  bins of $M_{group}$.  This result  is not necessarily
inconsistent  with  the trend  of  $``a''$  with  $\Sigma_N$ shown  in
Fig.~\ref{fig:FP_DENS}  (upper  panel),  as  binning the  sample  with
respect to  both $M_{group}$ and $\Sigma_N$ decreases  the sample size
in  each bin,  possibly washing  out the  ($2.5~\sigma$) environmental
trend in  Fig.~\ref{fig:FP_DENS}.  Fig.~\ref{fig:FP_MASS_DENS_b} shows
that $``b''$ does not depend  significantly on group mass for low mass
groups, while  it increases significantly with $\log  \Sigma_N$ in the
high  mass regime.  In the  bin  of highest  $M_{group}$ ($5.8  \times
10^{14} M_\odot$),  the slope  of the $``b''$--$\log  \Sigma_N$ linear
fit is significantly  positive, at more than $4~\sigma$,  for the case
where no correction  is applied, and at more  than $2.5 ~\sigma$, when
the  correction is  performed.   We conclude  that  the dependence  of
$``b''$ on local density  (Fig.~\ref{fig:FP_DENS}) is mainly driven by
galaxies residing in most massive clusters.

\subsection{The slopes of the FP from g through K}
\label{sec:slopes_optNIR}
{ The environmental  analysis of the waveband dependence  of the FP
  involves essentially two aspects.   We derive the relative variation
  of FP slopes, from $g$ through $K$, and compare such variation among
  different environmental  bins.  Since  the waveband variation  of FP
  slopes informs  on how stellar population properties  (i.e.  age and
  metallicity) vary with galaxy mass  (see paper II for details), this
  allows  us  to analyse  how  such  stellar  population variation  is
  affected by  the environment where galaxies  reside.  Moreover, as
  the  NIR light  closely  follows the  stellar  mass distribution  in
  galaxies and the contribution of  stellar populations to the tilt of
  the  FP in  K-band  vanishes (see  paper  II), the  NIR (K-band)  FP
  informs on  how the ratio of  dynamical to stellar  mass (and/or the
  non-homology)  of ETGs  vary with  mass, and  how this  variation is
  related  to  the  environment.   The  results  presented  below  are
  discussed, with regard to both aspects, in Sec.~10.2. }

In order to  derive the FP coefficients in  different environments, we
use the optical+NIR  sample of ETGs, binning galaxies  with respect to
different environmental properties, and deriving the FP slopes in each
bin.  In all cases, the  binning is performed as for the environmental
analysis    of   the    FP    intercept   from    $g$   through    $K$
(Sec.~\ref{sec:offset_optNIR}).  We correct the FP slopes for the same
effects mentioned  in Sec.~\ref{sec:slopes_rband}, i.e.  environmental
differences in  the average mass-to-light  and luminosity segregation.
To this effect,  we follow the same recipes as for  the analysis of FP
slopes in r-band (Sec.~\ref{sec:slopes_rband}).

\begin{figure*}
\begin{center}
\includegraphics[height=80mm]{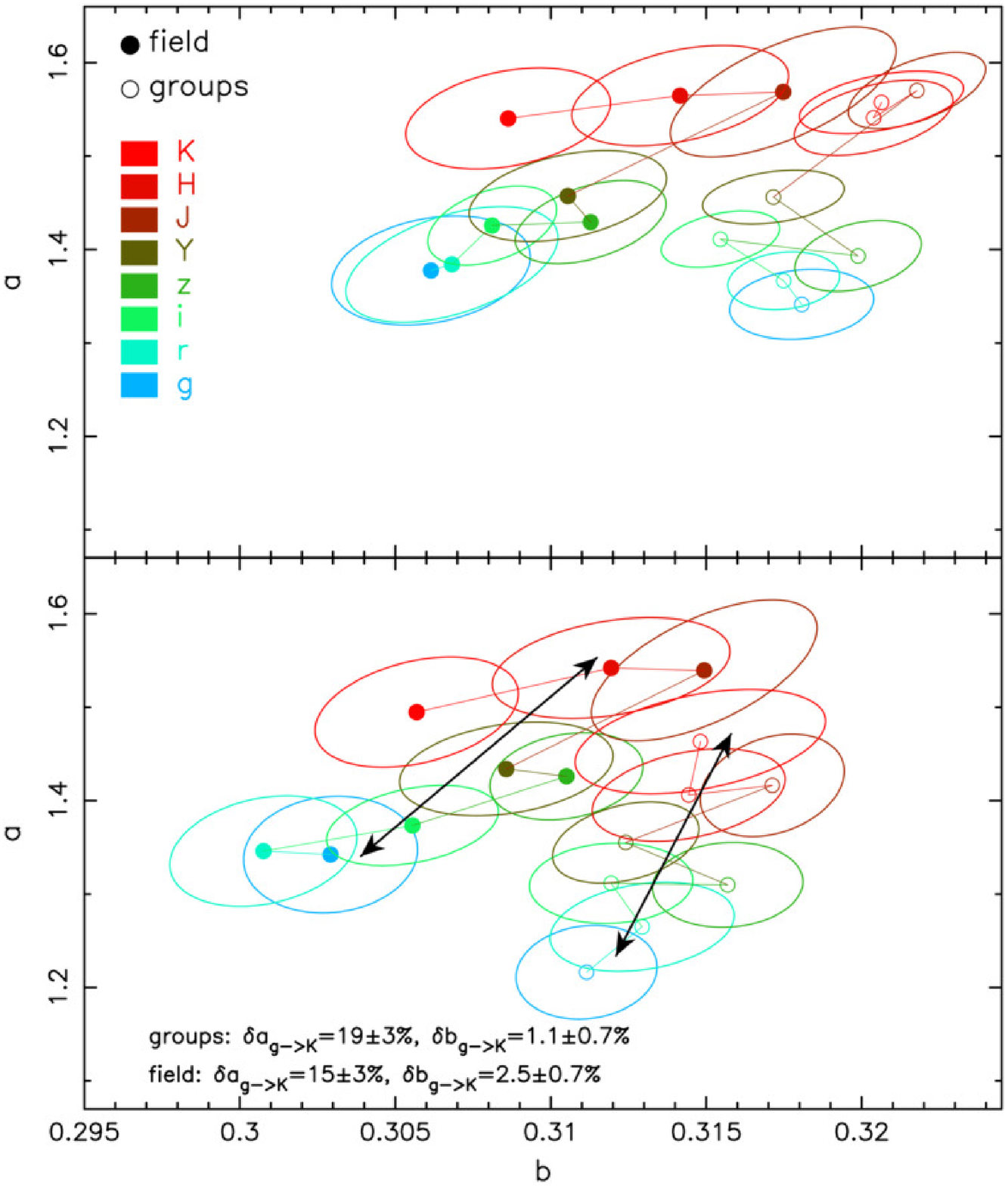}
\caption{Slopes $``a''$ and $``b''$ of the FP from $g$ through $K$ for
  galaxies residing in groups (empty circles) and in the field (filled
  circles).  Different  colours correspond to  different wavebands, as
  shown in the  lower-left corner of the upper  plot.  Ellipses denote
  $1~\sigma$ confidence  contours on  $``a''$ and $``b''$.   The upper
  and lower panels  correspond to the cases where  the FP coefficients
  are  either corrected  or  not  for the  difference  in the  average
  mass-to-light ratio  of galaxies  in the different  environments and
  the luminosity  segregation effect.  The  black arrows in  the lower
  panel plot the amount of  relative variation from $g$ through $K$ of
  the  FP slopes for  the two  subsamples, as  obtained by  assuming a
  power-law relation  between the $``a''$ ($``b''$)  and the effective
  wavelength  of  a  given  passband  (see the  text).   The  relative
  variations, \dagk \, and \dbgk, are reported in the lower--left part
  of the same panel.
\label{fig:FP_slopes_grizYJHK_field_group}
}
\end{center}
\end{figure*}

Fig.~\ref{fig:FP_slopes_grizYJHK_field_group} plots  the slopes of the
FP from  $g$ through $K$  for galaxies residing  in groups and  in the
field. The optical+NIR samples of  group and field galaxies consist of
$2,185$  and $1,359$ galaxies,  respectively. The  upper panel  of the
Figure  shows  the FP  slopes  for these  samples:  we  only detect  a
significant difference  in the  value of $b$  between field  and group
galaxies,   with   the   latter    having   lower   $``b''$   in   all
wavebands. However, to perform a proper comparison, we have to account
for differences in mass-to-light  ratio and luminosity segregation. To
this  effect, we  proceed as  for the  comparison of  field  and group
galaxies   in  the  r   band  (see   Sec.~\ref{sec:slopes_rband}),  by
extracting two  subsamples of galaxies with the  same distributions in
the space of structural parameters.  This selection leads to field and
group  sub-samples with  $1,264$ galaxies  each. The  corresponding FP
slopes,  from  $g$  through  $K$  are  shown in  the  lower  panel  of
Fig.~\ref{fig:FP_slopes_grizYJHK_field_group}.   We  are  now able  to
detect also a difference in the coefficient $``a''$. The $``a''$ turns
out  to be  smaller  for group,  relative  to field,  galaxies in  all
wavebands,  in  agreement  with  the  result for  the  r-band  samples
(Fig.~\ref{fig:FP_slopes_field_group}).   In   order  to  measure  the
variation  of FP  slopes  from $g$  through  $K$, we  assume a  linear
relation between  the $``a''$  ($``b''$) and the  effective wavelength
($\lambda$) of each waveband, i.e.
\begin{eqnarray}
a(\lambda) & = & p_a + q_a \log \lambda \label{eq:al} \\ 
b(\lambda) & = & p_b + q_b \log \lambda, \label{eq:bl} 
\end{eqnarray}
where $p_a$ ($p_b$)  and $q_a$ ($q_b$) are the  intercept and slope of
the linear relation.  The values  of $p_a$ and $q_a$ ($p_b$ and $q_b$)
are  derived by  an ordinary  least-squares fit  of  $``a''$ ($``b''$)
versus $\log  \lambda$.  The uncertainties  on $p_a$ and  $q_a$ ($p_b$
and $q_b$) are estimated from those on $``a''$ and $``b''$, accounting
for   the   correlated   errors   of   FP   slopes   among   different
wavebands~\footnote{Since  we  analyse the  same  sample  of ETGs  and
  velocity dispersions  are the same  in all different  wavebands, the
  uncertainties  on  $``a''$ and  $``b''$  from  $g$  through $K$  are
  correlated. To  estimate the  errors on $p_a$  and $q_a$  ($p_b$ and
  $q_b$), we randomly  shift the values of $``a''$  ($``b''$) from $g$
  through $K$,  and repeat  the fitting.  The  shifts are  computed by
  taking  into account the  correlation of  uncertainties. Had  we not
  accounted for  it, the errors on  $p_a$ and $q_a$  ($p_b$ and $q_b$)
  would have  been significantly overestimated, by almost  a factor of
  two.}.  We find that, for all the subsamples of ETGs analysed in the
present paper,  Eqs.~\ref{eq:al} and~\ref{eq:bl} are  able to describe
the trends of $``a''$ and $``b''$  with waveband with an accuracy of a
few percent.  They provide simple, empirical relations to quantify the
waveband variation of FP slopes.  For both field and group ETGs (lower
panel   of   Fig.~\ref{fig:FP_slopes_grizYJHK_field_group}),  we   use
Eqs.~\ref{eq:al} and~\ref{eq:bl}  to calculate the  relative variation
of $``a''$ and $``b''$ from $g$ through $K$, i.e.
\begin{eqnarray}
\delta a_{g->K} & = & \frac{a(\lambda_K)-a(\lambda_g)}{a(\lambda_g)} \label{eq:dal} \\ 
\delta b_{g->K} & = & \frac{b(\lambda_K)-b(\lambda_g)}{b(\lambda_g)} \label{eq:dbl} 
\end{eqnarray}
where $\lambda_g$ and $\lambda_K$ are the effective wavelengths of the
$g$ and $K$ passbands, respectively.  The values of \dagk \, and \dbgk
\,  define  a  segment  in  the  $``a''$--$``b''$  plane,  whose  size
quantifies the  amount of variation of  FP slopes from  the optical to
NIR. The segments  corresponding to the subsamples of  field and group
ETGs, as well  as the values of  \dagk \, and \dbgk, are  shown in the
lower  panel  of  Fig.~\ref{fig:FP_slopes_grizYJHK_field_group}.   The
uncertainty on  \dagk \, (\dbgk) is  obtained from those  on $p_a$ and
$q_a$ ($p_b$  and $q_b$).  In  general, the variations of  $``a''$ and
$``b''$ from  $g$ through $K$ are  similar for both  samples.  For the
$``b''$, field  galaxies tend to have a  marginally steeper variation,
with  \dbgk$=2.5\pm0.7\%$, compared  to \dbgk$=1.1\pm0.7\%$  for group
galaxies.  More in  general, one can notice that  while the $``a''$ is
always smaller for group relative to field galaxies, the difference in
$``b''$  between the  two samples  tends  to vanish  when moving  from
$``g''$ to $``K''$  (Sec.~\ref{sec:slopes_rband}). To further quantify
these  differences,  we  take  the   average  of  the  g-  and  r-band
coefficients,  and compare  them to  the  mean values  of $``a''$  and
$``b''$ in  the $JHK$ (NIR)  wavebands. For the field  (group) sample,
the $``b''$  changes from $0.302  \pm 0.003$ ($0.312\pm0.002$)  in the
optical to $0.311 \pm 0.003$ ($0.316 \pm 0.003$) in the NIR. Thus, the
difference  in  $``b''$  between   the  field  and  group  samples  is
significantly in  the optical, while  becomes consistent with  zero in
the NIR.  For the field (group) sample, the $``a''$ changes from $1.35
\pm 0.06$ ($1.23\pm0.04$) in the optical to $1.53 \pm 0.06$ ($1.42 \pm
0.05$) in the  NIR. Thus, the difference in  $``a''$ between the field
and  group samples is  essentially independent  of waveband,  with the
$``a''$ being smaller (at $\sim  2~\sigma$), for the group relative to
field sample in both the optical and NIR.

\begin{figure}
\begin{center}
\includegraphics[height=80mm]{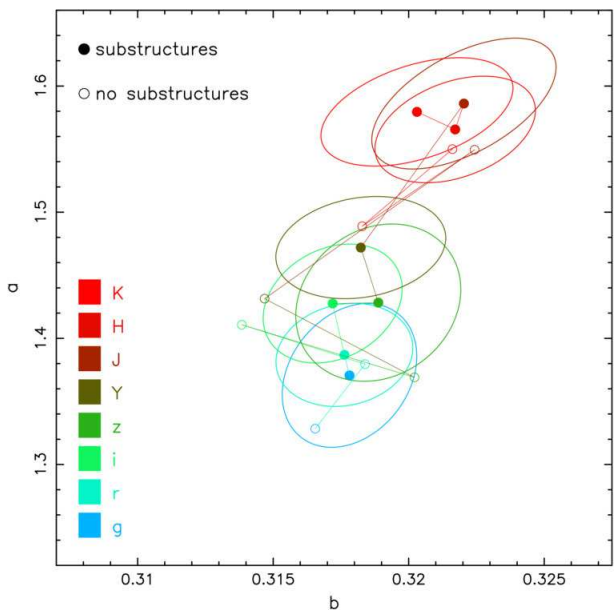}
\caption{The      same      as       the      lower      panel      of
  Fig.~\ref{fig:FP_slopes_grizYJHK_field_group}   but   for   galaxies
  residing in groups with (filled circles) and without (empty circles)
  substructures.
\label{fig:FP_slopes_grizYJHK_substr}
}
\end{center}
\end{figure}

Fig.~\ref{fig:FP_slopes_grizYJHK_substr} plots the  FP slopes from $g$
through  $K$  for  galaxies   residing  in  groups  with  and  without
substructures.   We do not  find any  significant difference,  for all
wavebands, between the two samples.

\begin{figure*}
\begin{center}
\includegraphics[height=100mm]{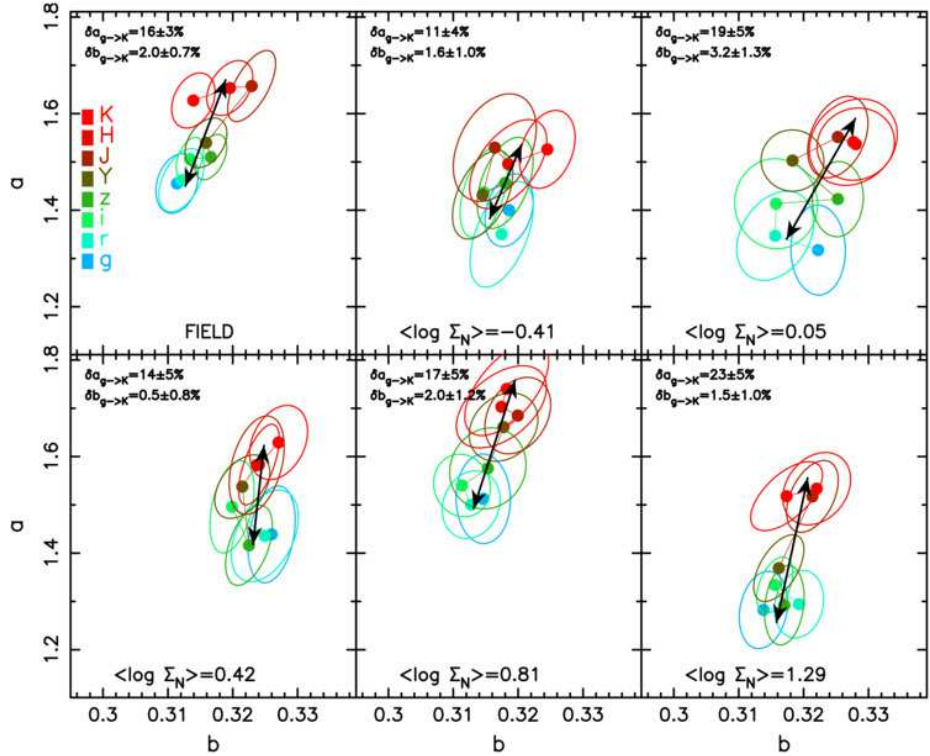}
\caption{Slopes $``a''$  and $``b''$  of the FP  in different  bins of
  local galaxy  density.  The  average value of  $\log \Sigma_N$  in a
  given bin is reported in  the lower part of the corresponding panel,
  increasing from  left to  right and top  to bottom.   The upper-left
  panel shows the results obtained for field galaxies. Ellipses denote
  the  $1~\sigma$ confidence  contours.  For each  panel, the  segment
  shows the  amount of  variation of FP  slopes from $g$  through $K$,
  obtained by a power-law fit of  $``a''$ and $``b''$ as a function of
  the effective wavelength of  each waveband. The amount of variations
  of  $``a''$ and $``b''$,  \dagk \,  and \dbgk,  are reported  in the
  upper-left corner of each panel.
\label{fig:FP_slopes_grizYJHK_dens}
}
\end{center}
\end{figure*}

\begin{figure*}
\begin{center}
\includegraphics[height=100mm]{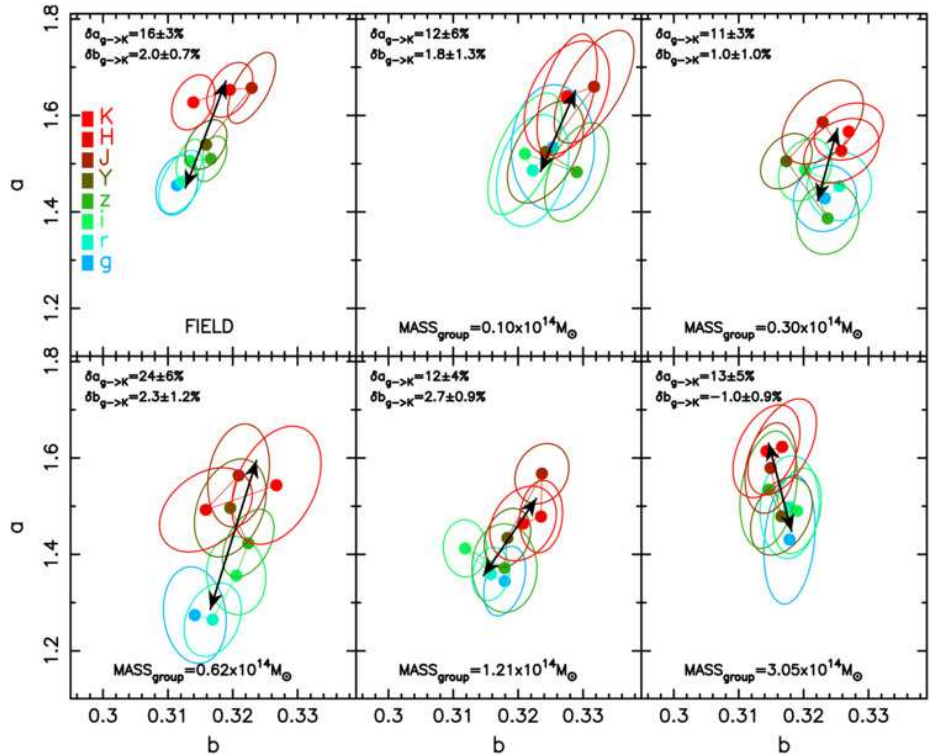}
\caption{The  same as  Fig.~\ref{fig:FP_slopes_grizYJHK_dens}  but for
  different bins of mass of parent galaxy groups.
\label{fig:FP_slopes_grizYJHK_mass}
}
\end{center}
\end{figure*}

Figs.~\ref{fig:FP_slopes_grizYJHK_dens}
and~\ref{fig:FP_slopes_grizYJHK_mass} plot  the slopes of  the FP from
$g$ through $K$ in different bins of local galaxy density, and mass of
the parent galaxy groups. In each environmental bin, we apply suitable
correction  factors to  the  slope  values, in  order  to account  for
environmental differences in  the average mass-to-light and luminosity
segregation. The  corrections are  computed as in  the analysis  of FP
slopes in r-band (Sec.~\ref{sec:slopes_rband}).  For all wavebands, we
apply the same corrections. This is  justified by the fact that (a) we
always analyse the same sample of ETGs in the different wavebands, and
(b)   as   shown  in   Sec.~\ref{sec:offset_optNIR}   the  amount   of
environmental  variation of  the  FP  offset is  very  similar in  all
wavebands, implying  similar differences in  the corresponding average
mass-to-light ratios.  We find that  the waveband dependence of the FP
is very similar in all  wavebands.  The $``a''$ tends to increase with
wavelength in  all environments. We  measure the amount  of end-to-end
variation  of $``a''$  and  $``b''$, across  the available  wavelength
baseline, by computing the quantities  \dagk \, and \dbgk (see above).
The value of \dagk \, spans the range $\sim 11\%$ to $\sim 23\%$, with
a  weighted-mean  value  of  $16\%$.   All  values  of  \dagk  \,  are
consistent within  the errors with that  of $16 \pm  3\%$ measured for
the  field sample.   The $``b''$  does not  change  significantly with
waveband,  with  a  marginal   tendency  to  increase  (\dbgk$>0$)  by
$1$--$3\%$ from $g$  through $K$ in all the  environments. Notice that
the  quantity \dbgk  \,  is positive  in  all bins  of $\Sigma_N$  and
$M_{group}$ with  the exception of the  the bin with  highest value of
parent group  mass ($M_{group}=3.05 \times 10^{14}  \, M_{\odot}$, see
Fig.~\ref{fig:FP_slopes_grizYJHK_mass}).  This is consistent with what
shown   in  Fig.~\ref{fig:FP_slopes_grizYJHK_field_group},   i.e.  the
$``b''$ increases by a few percent with waveband for the field sample,
while       it      is       constant       for      galaxies       in
groups.  Fig.~\ref{fig:FP_slopes_grizYJHK_mass} would also  imply that
this  different  behaviour  is  due  to galaxies  residing  in  massive
clusters.  The plots of  $``a''$ and  $``b''$ as  a function  of other
environmental quantities, such as the cluster-centric distance and the
$``global''$ mass surface density of  the parent galaxy groups, do not
add further  information to the  conclusions above, and are  not shown
here for brevity reasons.

To summarise,  we find  that the tilt  of the  FP, as measured  by the
coefficient $``a''$,  is different  between field and  group galaxies,
with  group   galaxies  having  lower  $``a''$.    The  difference  is
independent of the passband where the galaxy structural parameters are
measured. For the $``b''$, its variation with waveband is smaller than
$\sim  3\%$. However, in  this case,  an important  difference between
field  and group galaxies  emerges, i.e.   the $``b''$  increases with
waveband   (by   $2.5   \%$,   see   the  value   of   \dbgk   \,   in
Fig.~\ref{fig:FP_slopes_grizYJHK_field_group}),    while   for   group
galaxies the  variation with waveband is insignificant.  This seems to
be  especially true  for galaxies  residing in  most  massive clusters
(Fig.~\ref{fig:FP_slopes_grizYJHK_mass}),   for  which  \dbgk   \,  is
negative.  The  implications  of   these  findings  are  discussed  in
Sec.~\ref{sec:conc}.

\section{Discussion}
\label{sec:conc}
In the present work, we  have analysed the environmental dependence of
the FP coefficients on the waveband at which structural parameters are
measured. On what follows,  we discuss the environmental dependence of
the FP in the optical (Sec. 10.1) and using the combined optical$+$NIR
(Sec. 10.2).  In  Sec. 10.3 we examine the stellar  content of ETGs in
different environments. The most remarkable result we find here is the
variation  of  the  FP  offset   with  local  density.   As  shown  in
Sec.~\ref{sec:offset_sp},  this trend sets  strong constraints  on the
variation  of   stellar  population   properties  of  ETGs   with  the
environment.  This point is discussed in Sec.~\ref{sec:env_sp}.

\subsection{Environmental dependence of the optical FP relation}
\label{sec:conf_previous_works}

The first attempt to measure systematic differences in the FP relation
among  different  environments  has  been  performed  by~\citet{dCD92}
(hereafter   dCD92;   see   also~\citealt{DdC88}),  who   derived   FP
coefficients in the optical for  a sample of $31$ cluster (mostly Coma
and    Virgo)     and    $34$    $``field''$     galaxies    in    the
low-redshift-Universe. Field  galaxies included also  objects residing
in poor and loose groups. The authors found that the offset $``c''$ of
the  FP for  field galaxies  is significantly  different from  that of
galaxies in clusters.   They interpreted qualitatively this difference
as field galaxies being younger  than those in clusters, consistent to
what we  find here. However,  a direct and quantitative  comparison of
the  values of $``c''$  between our  study and  dCD92 is  not possible
because of  the different definition of environment  and the different
techniques   used  to  estimate   the  FP   intercept.   Due   to  the
uncertainties on FP coefficients, dCD92 did not detect any significant
variation of FP slopes with environment.

The first systematic study of  the FP relation in clusters, at optical
wavebands,  was performed  by~\citet{JFK96}  (hereafter JFK96).   They
derived the FP coefficients for ETGs, in Gunn-r, residing in 11 nearby
clusters (220 galaxies).   They found no evidence for  variation of FP
coefficients  with global cluster  properties like  richness, velocity
dispersion and  temperature of the  intra-cluster gas. Also,  they did
not detect  any correlation  between residuals around  the FP  and the
local environment,  characterized as the  cluster-centric distance and
the       projected      cluster       surface       density      (see
also~\citealt{LAB00}). These results seem  to contrast to what we find
here,  i.e.  that  both  $``c''$  and (to  minor  extent) $``a''$  and
$``b''$   depend   on   the   environment   (Secs.~\ref{sec:offset_fp}
and~\ref{sec:slopes_fp}, respectively).  However,  one has to consider
that  the number  of galaxies  per cluster  in JFK96  is  much smaller
(larger uncertainties)  than that in  each local density  bin analysed
here.   For instance, looking  at tab.~3  of JFK96,  one can  see that
their typical error on $``b''$ varies from $\sim 0.02$ to $\sim 0.06$,
among the different clusters, while  in our case this error amounts on
average  to $\sim$0.006.   The environmental  variation of  $``b''$ we
find ranges from $\sim$0.02 to $\sim 0.04$ (see the panel representing
the more massive systems in Fig.~\ref{fig:FP_MASS_DENS_b}), comparable
to the quoted uncertainties on $``b''$ in JFK96.  This explains why we
are able to detect such a small difference while JFK96 cannot.

\citet{BERN:06} (hereafter B06) analysed the offset of the FP relation
of ETGs in  low and high density regions (bins),  using a large sample
of ETGs from  SDSS.  The authors defined the  environment using the C4
cluster   catalogue   (based   on  SDSS-DR2;   see~\citealt{MILLER:05}).
Galaxies were assigned to the high  density bin using a cut in angular
separation  and  radial  velocity  around each  cluster,  rather  than
assigning    membership   to    each    object   as    we   do    here
(Sec.~\ref{sec:environment}).   Galaxies in the  low density  bin were
defined  using  a  cutoff  in  the  maximum  allowed  distance  for  a
$``field''$ galaxy  to the n-th  companion galaxy. This  criterion was
introduced by  B06 in order to  remove the contamination  in the field
sample  from galaxies  residing in  poor and  loose groups,  which are
likely  not included  in the  C4 catalogue.   On the  contrary,  the FoF
catalogue we  use here also includes  poor galaxy groups,  and we define
the  field sample  according to  the 3D  distance of  galaxies  to the
nearest group. Despite of  these differences and the different sources
of structural  parameters used to derive  the FP in  the present study
(i.e.  2DPHOT,  see paper I) and  B06 (i.e. the  SDSS Photo pipeline),
the  difference of  the FP  intercept among  field and  group galaxies
between our  study and that  of B06 is  very consistent. In  fact, the
latter  find that  the average  difference of  \mie \,  in r  band, as
derived from  the FP,  amounts to $0.075\pm  0.008$, with  galaxies in
denser   regions  having   fainter  surface   brightness   than  their
counterparts in the field. On the other  hand, we find $- \delta c / b
=  0.077 \pm  0.007$~mag (see  Sec.~\ref{sec:slopes_rband}),  which is
fully consistent with  the B06 result. We address  the implications of
this result in Sec. 10.3.

A systematic study of the environmental effects on FP coefficients has
been  recently  performed,   in  the  optical,  by~\citet{Donofrio:08}
(hereafter     D08),    in    the     framework    of     the    WINGS
project~\citep{Fasano:06}.   The  sample of  D08  consists of  $1,550$
early-type    galaxies    in    59    massive   clusters,    in    the
low-redshift-Universe ($0.04<z<0.07$).  Hence,  they analyse a similar
redshift range to  that considered by our study,  but with a different
range  in mass  of  the  parent systems  where  galaxies reside.   D08
detected  no  significant variation  of  FP  coefficients with  global
cluster  properties,   such  as  richness,   $R_{200}$,  and  velocity
dispersion.  On the contrary, we find evidence that the offset $``c''$
and  (to   minor  extent)   the  slope  $``b''$   of  the   FP  change
systematically  with the  global environment,  with  $``b''$ ($``c''$)
increasing  (decreasing)  with  respect  to  the  parent  group  mass,
$M_{group}$ (see Figs.~\ref{fig:offset_r_mass} and~\ref{fig:FP_MASS}).
We notice  that these trends  are probably detectable only  because of
the larger  range in $M_{group}$ spanned by  our group/cluster catalogue
(see Fig.   3) in  comparison with theirs  (400 $<$ $\sigma$  $<$ 1400
km/s,  with  $<\sigma>$  =  800  km/s).   

Moreover,  as discussed  in Sec.~\ref{sec:offset_fp},  the  trend with
global environment can  be entirely explained by the  dependence of FP
coefficients on  local density.   D08 found a  strong variation  of FP
coefficients  with the  local environment.   Both $``a''$  and $``b''$
were found to increase, and $``c''$ was found to decrease with respect
to the  (projected) local galaxy  density.  The trends of  $``b''$ and
$``c''$ from D08 are qualitatively  consistent with those we find here
(see     Secs.~\ref{sec:offset_rband}     and~\ref{sec:slopes_rband}).
However, after  correcting for systematic effects (see  red circles in
Fig.~\ref{fig:FP_DENS}), we  detect a $2.5\sigma$  indication that the
$``a''$ {\it decreases} from the low to the higher density regions, in
contrast with  D08. Considering ETGs  in the most massive  groups (see
Fig.~\ref{fig:FP_MASS_DENS_a}),  we  still  do  not  find  any  strong
variation with  local environment, as in D08.   A possible explanation
for this  discrepancy can be  the fact that  we have corrected  the FP
slopes  of   different  samples  of   ETGs  for  biases  due   to  the
environmental  dependence  of   the  average  mass-to-light  ratio  of
galaxies  and the  distribution  of  ETGs in  the  space of  effective
parameters, while  D08 did  not account for  either effects.   We also
notice that the  variation of $``b''$ with local  environment from D08
is much larger than  that detected here.  From Fig.~\ref{fig:FP_DENS},
we see that the end-to-end  variation of $``b''$ with respect to local
density amounts  to less  than $0.02$, while  D08 find a  variation of
$\sim 0.04$ (see their fig.~15).  This discrepancy can be explained by
the different  mass regime covered by  our cluster sample  and that of
D08, and by the fact that  we find the variation of $``b''$ with local
environment to depend on the mass of parent galaxy groups.  For groups
as  massive  as $\sim  5.8  \times 10^{14}  \,  M_\odot$,  we find  an
end-to-end variation of $``b''$ of  $\sim 0.04$ when none of the above
corrections is applied (in agreement with D08). This variation reduces
to $\sim 0.02$ after the corrections are performed.

\subsection{Environmental dependence of the FP from g through K}
\label{sec:env_dep_FP_gtoK}

For what concerns  the waveband variation of FP  coefficients, we find
that  the $``a''$  increases, by  $\sim 16\%$,  from $g$  through $K$,
independent of  the environment, while  the variation of  $``b''$ with
passband  is  smaller  than  $\sim  3\%$.  For  field  galaxies,  (see
Fig.~\ref{fig:FP_slopes_grizYJHK_field_group}), there  is a $3~\sigma$
evidence that the $``b''$ increases, by $\sim 2.5\%$, from the optical
to  NIR,  with the  difference  in  $``b''$  between field  and  group
galaxies  vanishing in  the NIR.   For group  galaxies,  in particular
those        residing       in       most        massive       systems
(Fig.~\ref{fig:FP_slopes_grizYJHK_mass}), no variation of $``b''$ with
waveband  is  detected.   As  discussed  in  paper  II,  the  waveband
dependence of FP slopes informs on the variation of stellar population
properties  with mass.   For the  entire SPIDER  sample,  the relative
variation of  $``a''$ and  $``b''$ from $g$  through $K$  implies that
ETGs, as  a whole, have  essentially coeval stellar  populations, with
metallicity  being larger  in more  massive galaxies.   As  shown from
eq.~7 of  paper II, a smaller value  of $``b''$ (as that  we find here
for field  galaxies) implies a steeper variation  of the mass-to-light
ratio  of  ETGs along  the  galaxy  (mass)  sequence. As  the  $``b''$
increases   from  the  optical   to  NIR   for  field   galaxies,  the
corresponding  mass-to-light ratio  variation  is related  to the  way
stellar population  properties change with mass.  In  other terms, the
environmental  variation of  $``b''$ might  imply that  group galaxies
exhibit  a shallower  relation between  stellar  population parameters
(i.e.   age  and  metallicity)   and  mass.   Interestingly,  this  is
consistent with the  recent findings of~\citet{Pasquali:10} (hereafter
PGF10), who found that  galaxies in denser environments have shallower
age--  and  metallicity--mass  relations.   This  is  also  a  natural
expectation of hierarchical models of galaxy formation, as galaxies in
denser  environments are those  accreted in  the group  environment at
earlier times,  having their  star formation quenched  earlier through
environmental-driven   effects  (e.g.    strangulation),   implying  a
shallower relation between (luminosity-weighted) age and stellar mass.
A flatter metallicity--mass relation  in denser environment might also
result from  the effect of tidal  stripping on the  more bound stellar
material   in   galaxies   (e.g.~\citealt{Kli09})  as   discussed   by
PGF10.\\ On  the other hand,  the difference in $``a''$  between field
and   group  galaxies   is  the   same   at  both   optical  and   NIR
wavebands. Group  galaxies have a larger  FP tilt, as  measured by the
coefficient $``a''$, likely  because of a waveband-independent origin.
A  possible  explanation  is  a  different  variation  of  dark-matter
fraction and/or  non-homology along the mass sequence  of ETGs between
galaxies in different environments. For instance, group galaxies might
have a larger  tilt because of a steeper  relation between dark-matter
fraction  and  mass, due  to  (i)  tidal  interactions between  galaxy
dark-matter  halos and parent  group halos,  i.e.  lower,  relative to
higher, mass galaxies  are stripped more of their  dark-matter halo as
they fall into the cluster  potential well; (ii) central galaxies have
higher dark-matter fractions, as  they acquire part of the dark-matter
halo of  the in-falling satellites.   The first scenario  might result
from  the fact that  the total  time span,  during which  galaxies are
stripped,  is   larger  for  low   mass  galaxies  (see   e.g.   eq.~9
of~\citealt{Cattaneo10}). However, tidal  forces are stronger for more
massive galaxies,  as a  short orbital  decay time for  $\rm L  > L^*$
galaxies  is expected~\citep{Barnes89}.   The scenario  (ii)  might be
more plausible, provided that  stripped galaxies have mostly late-type
morphology.   This would not  change significantly  the peak  value of
dynamical mass of  ETGs as a function of  environment, consistent with
Fig.~\ref{fig:mdyn_r_dens}, and increase only the dark-matter fraction
for higher mass  systems, implying a smaller tilt  of the FP relation.
{ Notice that a steeper  relation between stellar to dynamical mass
  ratio and  mass in ETGs for  group (relative to  field) galaxies has
  recently been suggested also  from~\citet{SB:09}, as a result of the
  wider  range  of  luminosity-weighted  ages  characterizing  cluster
  relative to field galaxies.}   These points will be further analysed
in  a forthcoming  contribution  in this  series,  using estimates  of
dynamical and stellar mass for  ETGs and analysing their dependence on
environment.

  It is interesting  to notice that the NIR  FP, and its environmental
  dependence, can be connected to the stellar mass FP recently derived
  by~\citet{HB:09}.   The stellar  mass  FP is  obtained by  replacing
  luminosity (i.e. surface brightness) in the FP equation with stellar
  mass.   Since the  K-band  light closely  follows  the stellar  mass
  distribution in galaxies and the contribution of stellar populations
  to the tilt of the FP vanishes in the NIR (see paper II), the K-band
  FP should essentially  coincide with the stellar mass  FP.  In fact,
  \citet{HB:09} report a value of $a \sim 1.54\pm0.02$ for the stellar
  mass FP,  while we measure $a  \sim 1.55\pm0.02$ for  the K-band FP.
  This  reinforces the fact  that the  environmental difference  of FP
  coefficients   we  detect   in  the   NIR  bands,   among  different
  environments (see above), implies an environmental difference in the
  relation between  the ratio of dynamical-to-stellar  mass and galaxy
  mass. Given the  excellent agreement of the K-band  and stellar mass
  FP's,  one  might actually  use  either  relations  to derive  ETG's
  stellar  masses,  and test  how  realistic  are  the estimates  from
  stellar population models.

\subsection{Stellar content of ETGs in different environments}
\label{sec:env_sp}
The  environmental variation  of the  FP intercept,  `$``c''$, implies
that the  average mass-to-light  ratio of ETGs  increases from  low to
high density regions.  { As seen in Eq.~\ref{eq:age_met_dens}, this
  variation can be interpreted as  a difference in the stellar content
  of  ETGs, at fixed  mass, provided  that the  ratio of  dynamical to
  stellar  mass, $M/M_{\star}$,  does not  increase, on  average, with
  local  density.   There  are   several  arguments  against  such  an
  increase.   First, Fig.~\ref{fig:mdyn_r_dens}  shows  that the  peak
  value  of  the quantity  $\sigma_0^2  \times  R_e$  -- a  proxy  for
  dynamical mass  -- does  not change with  the environment.   So, the
  $M/M_{\star}$ can increase with local density, only if stellar mass,
  at fixed  $M$, decreases with $\Sigma_N$.  However,  as galaxies are
  accreted into groups and clusters,  they are expected to be stripped
  off of their dark-matter halos, resulting, eventually, into a larger
  $M_{\star}$, for  fixed $M$, at high density.   These arguments make
  plausible our  assumption of constant $M/M_{\star}$,  that allows us
  to use the  offset of the FP  as a tool to analyse  the variation of
  stellar population  properties as a  function of environment.  It is
  interesting to note that,  as shown in fig.~8 of~\citet{BERN:06}, at
  fixed   luminosity,  $L$,   dynamical  mass   does  not   depend  on
  environment,  while the  $g-r$ colour  (which  is a  good proxy  for
  $M_{\star}/L$)  does.    This  implies  that,  at   fixed  $L$,  the
  $M/M_{\star}$ should change with environment. We notice that this is
  not   inconsistent   with   the   above   assumption   of   constant
  $M/M_{\star}$, as we are assuming that the $M/M_{\star}$ is constant
  at fixed $M$ (not $L$)}.

We find a significant difference of  the FP offset, in r band, between
field and group galaxies, fully consistent with that measured from B06
(see  previous  section).  Using  spectral  abundance indicators,  B06
concluded that the difference in $``c''$ is due to a difference in age
between field and  cluster ETGs, with field galaxies  being younger by
$\sim  1~Gyr$. Using  the  information  encoded in  the  FP offset  at
different wavebands,  we show that  the variation of the  FP intercept
with  local   galaxy  density  implies  a   variation  in  logarithmic
(luminosity-weighted) age per decade  of local density of $\delta \log
t/\delta \log  \Sigma_N =0.048 \pm  0.006$.  From Eq.~\ref{eq:c_dens},
taking into account the difference  of $``c''$ between field and group
galaxies (see  previous section), the  value of $\delta  \log t/\delta
\log \Sigma_N$ translates to a  relative difference in age of $\sim 18
\pm  2\%$ between  the field  and  group environments.   For a  galaxy
formation epoch  of $10~Gyr$, this implies an  age absolute difference
of $\delta t \sim 1.8\pm 0.2~Gyr$,  larger than what found by B06. The
finding  that  field ETGs  are  younger  than  those in  high  density
environments  is qualitatively consistent  with some  previous studies
(see  Sec.~\ref{sec:intro}).~\citet{Clemens:09} found that  field ETGs
are $\sim 2~Gyr$ younger than their cluster counterparts, in agreement
with the  value of  $\delta t$  we measure here.   The same  amount of
difference  in  age  was  also  found  by~\citet{Thomas:05}.   Similar
results, but based  on smaller samples of ETGs,  were also obtained by
(e.g.) \citet{Kunt:02, TF:02, Sanch:06, deLaRosa:07}. { Our results
  are also consistent with those of~\citet{Cooper:10}, who found that,
  at given  colour and luminosity  (and hence stellar  mass), galaxies
  with   older   stellar   populations   favour  regions   of   higher
  overdensity.}  We notice  that all these studies have  been based on
the analysis of spectral features (i.e. line indices) of ETGs.  Hence,
the  inferred differences  in age  refer to  the inner  galaxy region,
typically inside  one effective radius.   As already noticed  in paper
II, the information  provided by the waveband dependence  of the FP is
more related  to the  global properties of  the light  distribution in
galaxies (as measured by  the structural parameters).  In this regard,
it is  not affected by radial  population gradients in  ETGs and their
possible   dependence    on   environment~\citep{LaB:05}.    Recently,
\citet{Rogers:10} have performed a Principal Component Analysis of the
SDSS spectra of ETGs residing  in groups and clusters, spanning a wide
range in mass (similar to that of our FoF catalogue). They find galaxies
populating the lowest mass halos ($M_{group} \sim 10^{12} M_\odot$) to
be  younger, by  $\sim 1$~Gyr,  than those  in most  massive clusters.
Using the  linear fit  in Fig.~\ref{fig:offset_r_mass}, we  estimate a
variation  of the FP  offset, over  the entire  range of  parent group
mass, of $\sim -0.015$.  From the measured values of $b_{_{\Sigma_N}}$
and $\delta \log t  / \delta \Sigma_N$ (see Sec.~\ref{sec:offset_sp}),
this difference  translates to  a difference in  age of $\sim  12 \%$,
i.e.  $\sim 1.2$  Gyr (for a formation epoch  of $10$~Gyr), consistent
with~\citet{Rogers:10}.

Recently,   \citet{Thomas:09}  have   analysed  a   large   sample  of
morphologically  selected   ETGs  from  the   SDSS,  deriving  stellar
population     properties,     i.e.      age,     metallicity,     and
$\alpha$-enhancement,   for   galaxies   in  low-   and   high-density
environments. In  contrast to the  works listed above,  they concluded
that the luminosity-weighted age of the bulk of ETGs is independent of
the environment.   Low- and high-density  environments actually differ
because  of the  fraction of  galaxies showing  signs of  ongoing star
formation.  The  percentage  of  $``rejuvenated''$ ETGs  is  found  to
increase significantly  in the lower  density environments. It  is not
clear  if this  scenario is  also  able to  explain the  environmental
dependence of the FP.   As noticed in Sec.~\ref{sec:offset_rband}, the
variation of FP intercept in  r band remains essentially the same when
restricting  the  analysis  to   the  subsample  of  ETGs  with  lower
contamination from objects  with non-genuine ETG morphology. Moreover,
the FP offset is computed  from the median values of the distributions
of  effective parameters  and velocity  dispersions,  hence reflecting
more closely the bulk properties of the ETG's population.
 
Another  important finding  of the  present work  is the  lack  of any
evidence for ETGs in the field to be less metal-rich than those in the
cluster environment,  as found, for  instance, by~\citet{GALL:06}. The
optical+NIR  variation  of  the  FP  offset implies  a  difference  in
metallicity of $-0.043 \pm 0.023$  per decade in local galaxy density,
corresponding to  a difference  of $\sim 16  \pm 8\%$ between  ETGs in
low-  and  high-density  environments.   Hence,  our  FP  analysis  is
consistent  with field  galaxies being  more metal-rich  than  ETGs in
groups.   Rose et al.~(1994)  found a  similar result  and interpreted
this  difference  as  reflecting  the  fact that  star  formation  and
chemical enrichment  in ETGs  in clusters were  truncated at  an early
epoch, so ETGs were influenced by the environment at an early phase of
galaxy evolution. For comparison, ~\citet{BERN:06} found no detectable
difference in the  metallicity of field and cluster  ETGs, while other
studies,   e.g.~\citet{Thomas:05,   deLaRosa:07,   Clemens:09}   found
evidence for galaxies in dense environments to be less metal rich than
those in the field.

Different from most of the works  listed above, we do not only analyse
the  average difference  between  the properties  of  group and  field
galaxies,  but also  follow  the  variation of  FP  coefficients as  a
function of  local galaxy density. We  find that the offset  of the FP
smoothly changes  from the  highest density regions,  in the  cores of
parent  galaxy groups,  through  the outskirt  group  regions and  the
field.  Such variation is caused by a variation in age, i.e. ETGs have
progressively  younger  stellar  populations  from the  cluster  cores
through   the   field,   in   agreement  with   the   recent   results
of~\citet{Bernardi:09} and~PGF10,  who found that, for  a given parent
halo mass,  central galaxies  are systematically older  than satellite
galaxies. In contrast to~PGF10, we  find a positive (or eventually no)
gradient in  galaxy metallicity as  a function of local  density, that
would imply centrals to be  either less metal-rich or as metal-rich as
satellites  (see  above).   

Our  findings can  be compared  to the  expectations  of semi-analytic
models   (SAMs)  of  galaxy   formation~\citep{deLucia:06}  (hereafter
deL06).  The  model of deL06  predicts an age variation,  from cluster
cores to the field, of $\sim  2$~Gyr (see their fig.~8).  The trend is
mostly due  to the fact that,  in a hierarchical scenario,  halos in a
region of the Universe doomed to become a cluster are those collapsing
earlier  and  merging faster.  This  would  also  imply that  the  age
variation  as a function  of local  density might  be larger  for more
massive halos, in agreement with our finding that the variation of the
FP offset with  the local environment is larger  for galaxies residing
in clusters,  relative to groups.  Over the  large environmental range
spanned by the SPIDER sample  (three decades in local galaxy density),
we find  an age  variation of $\delta  \log t  = 0.14 \pm  0.02$, i.e.
$\delta t / t  \sim 32 \pm 5\%$ ($\sim 3.2 \pm  0.5$~Gyr, for a galaxy
formation epoch of  $10$~Gyr).  For a pure age  model of the variation
of  FP offset  with environment  (see  Sec.~\ref{sec:offset_sp}), this
difference reduces to $\delta  t / t \sim 28 \pm 3  \%$ ($\sim 2.8 \pm
0.3$~Gyr).  Both  values  of $\delta  t  /t  $  are larger  than  that
predicted  from  the  SAM,  but  not much  different  considering  all
assumptions  and   different  definitions  of   environments  in  both
cases.  We can  also notice  that SAMs  predict field  galaxies  to be
less-metal rich than their counterparts  in groups, in contrast to our
FP analysis.  As discussed by PGF10, current SAMs still do not provide
an  accurate description of  galaxy's metallicities:  several recipes,
such as the treatment  of SN feedback (see e.g.~\citealt{Bertone:07}),
tidal  interactions, and  recycling of  SN  ejecta, might  be able  to
improve the comparison  with the the environmental trends  seen in the
data.


\section{Summary}
\label{sec:summ}
We have  analysed the environmental  dependence of the FP  relation of
ETGs, combining  optical (SDSS) and NIR (UKIDSS)  data. Environment is
defined using the  largest group catalogue, based on  3D data, generated
from SDSS in the low-redshift-Universe  ($0.05 \le z \le 0.095$).  The
main results can be summarised as follows.
\begin{itemize}
 \item[$\bullet$]  The  intercept, $``c''$,  of  the FP  significantly
   changes with  environment. Local galaxy density is  the main driver
   of the  environmental dependence, with $``c''$  decreasing from low
   to high  density regions.   Weaker correlation is  observed between
   $``c''$  and the  mass and  mass density  of parent  galaxy groups.
   These latter  trends are  entirely explained by  the change  in the
   average  local density of  galaxies residing  in different  bins of
   (halo) mass  density.  The behaviour  of $``c''$ with  local density
   becomes steeper as the parent group mass increases.
 \item[$\bullet$] { The variation of $``c''$ with local environment
   implies that galaxies  in higher density regions have  on average a
   larger mass-to-light  ratio. Since the typical  dynamical mass does
   not change significantly with  environment and the trend of $''c''$
   is similar in all wavebands, from $g$ through $K$, we conclude that
   the difference in mass-to-light ratio is mainly due to a difference
   of  galaxy stellar population properties}.
 \item[$\bullet$]   The  variation   of  $``c''$   implies   that  the
   luminosity-weighted  age smoothly increases  with local  density by
   about $0.048$~dex ($\sim 11 \%$) per decade in local density, while
   metallicity  tends to  decrease.  Field  galaxies are,  on average,
   $18\%$ younger than their cluster counterparts.  As galaxies in our
   sample  span a range  of three  decades in  local density,  the age
   difference between galaxies in  the highest density cluster regions
   and the field  amounts to $\sim 32\%$. We have  also found (at $2.5
   \, \sigma$) that  the variation in age per  decade of local density
   might be  larger, up to a  factor of two, for  galaxies residing in
   more massive groups.
 \item[$\bullet$] The slope $``a''$ is smaller (at $\sim 3 \, \sigma$)
   for galaxies in groups compared  to the field ones. Consistent with
   that,  we also find  a weak  ($2.5 \,  \sigma$) correlation  in the
   sense that  $a$ decreases as the local  density increases.  $``b''$
   tends to  become higher when $\Sigma_N$ increases,  which is mainly
   due  to galaxies  inhabiting  more massive  clusters,  as for  poor
   groups  no trend is  detected.  The  difference in  $``a''$ between
   group and field galaxies seems to be similar in all wavebands, from
   $g$ through $K$,  maybe implying that the different  tilt of the FP
   between  group  and  field  galaxies  is  related  to  a  different
   variation  of dark-matter  fraction and/or  non-homology  along the
   mass sequence of ETGs.
 \item[$\bullet$] The waveband variation of $``a''$ is very similar in
   all  the  environments: it  increases,  by  $\sim  16\%$, from  $g$
   through $K$. $``b''$ varies by  less than $\sim 3\%$ with waveband.
   For field  galaxies, $``b''$  increases by $\sim  2.5 \%$  from $g$
   through $K$ (at $3~\sigma$), with the difference in $``b''$ between
   group and field galaxies becoming smaller in the NIR.  For galaxies
   in more massive parent halos, no variation of $``b''$ with passband
   is detected at  all. These findings may imply  that the correlation
   between  stellar   population  properties  and   mass  is  somewhat
   shallower for galaxies residing in higher density environments.
\end{itemize}
As a final  remark, we can look at  the FP of ETGs as a  tool to probe
physical  mechanisms  shaping  galaxies  as  we  observe  today.   The
environmental variation of its  coefficients may be interpreted in the
following way:  $``a''$ - reflects  some effect like homology  or dark
matter content;  $``b''$ is connected  to how the  stellar populations
vary with mass; and $``c''$  shows how the bulk of system's luminosity
varies.  These rules can be applied in the study of ETGs as a function
of environment, and at higher redshift, helping us to comprehend, even
if just in a phenomenological fashion, how galaxies form and evolve.

\section*{Acknowledgements}
We are  thankful to  Jason Pinkney for  making the  substructure codes
available.  We thank S.G.  Djorgovski, A.W.  Graham, I.  Ferreras, and
G. de  Lucia for  helpful comments and  suggestions.  { We  would also
  like to thank the anonymous referee for several useful suggestions.}
This research has  made use of the SAO/NASA  Astrophysics Data System,
and  the NASA/IPAC Extragalactic  Database (NED).   We have  used data
from the 4th data release of  the UKIDSS survey, which is described in
detail   in   \citet{War07}.     The   UKIDSS   project   is   defined
in~\citet{Law07}.   UKIDSS uses  the UKIRT  Wide Field  Camera (WFCAM;
Casali et al, 2007).  The photometric system is described in Hewett et
al  (2006),  and  the  calibration  is described  in  Hodgkin  et  al.
(2009). The  pipeline processing and science archive  are described in
Irwin et al (2009, in prep) and Hambly et al (2008).  UKIDSS data have
been  analyzed using  the Beowulf  system  at INAF-OAC~\citep{CGP:02}.
Funding for  the SDSS and SDSS-II  has been provided by  the Alfred P.
Sloan Foundation, the Participating Institutions, the National Science
Foundation, the  U.S.  Department of Energy,  the National Aeronautics
and Space Administration, the  Japanese Monbukagakusho, the Max Planck
Society, and  the Higher Education  Funding Council for  England.  The
SDSS Web  Site is  http://www.sdss.org/.  The SDSS  is managed  by the
Astrophysical Research Consortium  for the Participating Institutions.
The  Participating Institutions  are  the American  Museum of  Natural
History,  Astrophysical   Institute  Potsdam,  University   of  Basel,
University of  Cambridge, Case Western  Reserve University, University
of Chicago,  Drexel University,  Fermilab, the Institute  for Advanced
Study, the  Japan Participation  Group, Johns Hopkins  University, the
Joint  Institute for  Nuclear  Astrophysics, the  Kavli Institute  for
Particle Astrophysics  and Cosmology, the Korean  Scientist Group, the
Chinese Academy of Sciences  (LAMOST), Los Alamos National Laboratory,
the     Max-Planck-Institute     for     Astronomy     (MPIA),     the
Max-Planck-Institute   for  Astrophysics   (MPA),  New   Mexico  State
University,   Ohio  State   University,   University  of   Pittsburgh,
University  of  Portsmouth, Princeton  University,  the United  States
Naval Observatory, and the University of Washington.

\label{lastpage}

\end{document}